\documentclass[preprint,onecolumn,showkeyword
 amsmath,amssymb,
 aps,
]{revtex4-1}
\usepackage{graphicx,color}
\usepackage{dcolumn}
\usepackage{bm}

\begin{document}

\title{A Note on the "Various Atmospheres over Water Oceans on Terrestrial Planets" with a One-Dimensional Radiative-Convective Equilibrium Model}
\author{Tetsuya Hara}
\email{hara@cc.kyoto-su.ac.jp}
\affiliation{Department of Physics, Kyoto Sangyo University, Kyoto 603-8555, Japan}
\author{Anna Suzuki}
\email{i1785064@cc.kyoto-su.ac.jp}
\affiliation{Department of Physics, Kyoto Sangyo University, Kyoto 603-8555, Japan}
\author{Masayoshi Kiguchi}
\email{kiguchi-lab@abox3.so-net.ne.jp}
\affiliation{Kindai University Research Institute for Science and Technology, HigashiOsaka 577-8502, Japan}
\author{Akika Nakamichi}
\email{nakamichi@cc.kyoto-su.ac.jp}
\affiliation{Institute of General Education, Kyoto Sangyo University, Kyoto 603-8555, Japan}

\providecommand{\keywords}[1]{\textbf{\textit{Index terms---}} #1}


\begin{abstract}
It has been investigated the possibility of the various atmospheres over water oceans.  We have considered the H$_2$ atmosphere and He atmosphere concerning to N$_2$ atmosphere over oceans.  One of the main subjects in astrobiology is to estimate the habitable zone.  If there is an ocean on the planet with an atmosphere, there is an upper limit to the outgoing infrared radiation called the Komabayashi-Ingersoll limit (KI-limit). This limit depends on the components of the atmospheres. We have investigated this dependence under the simple model, using the one-dimensional gray radiative-convective equilibrium model adopted by Nakajima et al. (1992). 
The outgoing infrared radiation ($F_{IRout}$) with the surface temperature ($T_s$) has shown some peculiar behavior.  The examples for H$_2$, He, and N$_2$ background gas for H$_2$O vapour are investigated. 
There is another limit called the Simpson-Nakajima limit (SN-limit) mainly composed of vapour.  This steam limit does not depend on the background atmosphere components.  
 Under super-Earth case ($g=2\times$9.8 m/s$^2$), several cases are also calculated.  The KI-limit dependence on the initial pressure is presented.  The various emission rates by Koll \& Cronin (2019) are investigated.
\vspace{0.1cm}
{\flushleft{{\it Keywords:} Hycean Worlds $-$ Exoplanets atmospheres $-$ H$_2$ atmospheres $-$ He atmospheres $-$ Komabayashi-Ingersoll limit $-$ Simpson-Nakajima limit.}}
 \end{abstract}

\keywords{ }

\maketitle


\section{Introduction}

It has been pointed that hydrogen could play an important role in the early stages of terrestrial planet history, when the disk gas is almost hydrogen and He at first (Sekiya, Nakazawa, \& Hayashi 1981; Pierrehumbert \& Gaidos 2011; Koll \& Cronin 2019).   It has been made-up word 'Hycean' referred to H$_2$ rich atmospheres over the massive oceans (Madhusdhan et al. 2021).  
   As hydrogen is light, it is said to escape to outer space (Sekiya, Nakazawa \& Hayashi 1980), and it will take a long time to decrease to He dominant atmosphere (Hu et al. 2015).

Here we investigate cases of background gas with H$_2$ atmospheres over oceans. We also investigate cases with He atmospheres over oceans as well as N$_2$ atmospheres over oceans.  
It is taken the One-Dimensional Radiative-Convective Model adopted by Nakajima et al. (1992) for the gray atmosphere, which has been followed and investigated by many researchers (eg. Bressler \& Shaviv 2015).  The atmosphere is assumed to consist of the non-condensible  component (mainly N$_2$, or H$_2$, He) and condensible  component (eg. H$_2$O).   For the stratosphere, radiative equilibrium is assumed.  About the radiative transfer, the atmosphere is considered to be transparent to solar radiation in the optical range.  In the infrared range, the absorption coefficient is taken to be constant and independent of wavelength which is said to be a gray atmosphere. 
The radiation transfer equation is integrated by using the Eddington approximation.

There are two important limits such as "Komabayashi-Ingersoll limit" (KI-limit; Komabayashi 1967; Ingersoll 1969) which is called "dilute limit" by Koll \& Cronin (2019) and "Simpson-Nakajima limit" (SN-limit; Simpson 1927; Nakajima et al. 1992; Goldblatt et al. 2013) which is called "steam limit" by Koll \& Cronin (2019).  KI-limit is the upper limit for the emission from the top of the atmosphere in the Infrared radiation ($F_{IRtop})$ which depends on the component of the atmospheres and gravity.

SN-limit is the approximation limit of  $F_{IRtop}$ for H$_2$O atmosphere, which becomes the dominant component of the atmosphere in the high approximate temperature ($T \geq  500 \sim$ 600 K), so it does not depend on the component of the background atmosphere.   
SN-limit depends only on the gravity in this paper for $\simeq $ 294 W/m$^2$ for $g$= 9.8m/s$^2$ and $\simeq $340 W/m$^2$ for $g= 2\times $9.8 m/s$^2$, respectively.

For H$_2$ background atmospheres, it is characteristic that $F_{IRtop}$ increases with surface temperature $T_s$ at the first time, however, it seems to reach the upper limit, then decrease down and again increase to reach the second saturated limit denoted as the SN-limit (Goldblatt \& Watson 2012).  This feature is noticed by Suzuki (2017) and is named as "Souffl$\acute{e}$ Effect"  by Koll \& Cronin (2019).  One of the purpose of this work is to investigate the various limits and mechanism of these features.

We calculate cases of the surface accelerating gravity $g$=9.8 m/s$^2$ and 2 $\times$ 9.8 m/s$^2$.  The latter corresponds to almost ten times of Earth mass, called super-Earth (Madhusudhan et al. 2021).  Although it is pointed out that molecular hydrogen interacts strongly with infrared radiation via collision-induced absorption (CIA) in the high pressure situation (Wordsworth \& Pierrehumbert 2013), we have not considered such effects for H$_2$ and N$_2$ here.   We assume that background gases except H$_2$O are transparent.

  The only assumed different part among H$_2$, He and N$_2$ is the mean molecular weight $\bar{m}$ of the background gas, given by 
\vspace{-0.5cm}
  \begin{equation}
     \tilde{m}=m_n x_n+m_v x_v.
  \end{equation}
The parameter $m_n$ and $m_v$ are the molecular (or atomic) weight of the non-condensible and condensable (vapour) components, respectively.  And the parameters $x_n$ and $x_v$ are the mole fractions of the non-condensible  and condensible  components, respectively.  


In Sect. II, the method of the model is outlined.  It is applied to H$_2$ atmosphere in Sect. III and applied to He atmosphere in Sect. IV.   In Sect. V,  the KI-limit dependence on the initial pressure is investigated.  The results and discussion are deployed in Sect VI..

\section{One-Dimensional Radiative-Convective Model}
We have followed the method of Nakajima et al. (1992) for the One-Dimensional Radiative-Convective Model.  Under the tropopause, the adiabatic lapse rate is adopted where water vapour is saturated.  It is assumed that the saturation water vapour pressure  $p^{*}$ is derived under the Clausius-Clapeyron relationship and given by
  \begin{equation}
     p^*(T)=p^{*}_o \exp \left( -\frac{l}{RT} \right)  ,
  \end{equation}
where $ T$ and $R$ are temperature, the gas constant, respectively. 
The $l$, and $p^*_o$ are the latent heat of condensible component ($l=43655 $J mol$^{-1}$ ), and the constant for water saturation curve ($p^{*}_o=1.4 \times 10^{11} $ Pa).

    For the troposphere, it is assumed to be pseudoadiabatic lapse rate for the temperature gradient with pressure as
\begin{equation}
    \left(\frac{\partial T}{\partial p} \right)=\frac{\frac{RT}{pc_{pn}}+\frac{x_v^*}{x_n}\frac{l}{pc_{pn}}}{x_n+x_v^*\frac{c_{pv}}{c_{pn}}+\frac{x_v^*}{x_n}\frac{l^2}{RT^2c_{pn}}}  ,
\end{equation}
where  $c_{pv}$, and $c_{pn}$ are the mole specific heat at constant pressure of condensable and non-condensible components, respectively.

Nakajima et al. have assumed that the molecular weights are the same for condensible  and non-condensible substances for simplicity.  It is checked that the above equation (3) is applicable to the different molecular weights and the derivation of the above equation is commented in Appendix A.

   Given the non-condensible pressure $p_{n\ 0}$ and temperature at the surface $T_s$, $T(p)$ is obtained up to the tropopause of the atmosphere.  There appears to be a height where the net convergence becomes positive in the upper levels of the atmosphere. The position of the tropopause is taken to be there.

\subsection{Upper limit (KI-limit)}

One of the main problems in astrobiology is to find out the habitable zone where liquid water could exist.  The injection from the central emitting object (Sun) increases when the planet comes near to the center.  On the other hand, there is an upper limit that the planet could emit radiation.  
It depends on the atmospheric component and this paper investigates of the dependence under a simple model.


The optical depth $\tau$ is defined as
 \begin{equation}
  \tau =\int_{z}^{\infty } \kappa \rho dz .
\end{equation}
The only assumed different part of H$_2$ and He from N$_2$ is the mean molecular weight of the background gas.  The different point is the following equation about the optical depth $\tau $ (Nakajima et al. 1992) 
\begin{equation}
  d\tau=(\kappa _v x_v m_v +\kappa _n x_n m_n) \frac{dp}{\bar{m} g},
\end{equation}
where $p$ is the pressure and $g$ is the acceleration of gravity.  
As we assume $\kappa _n$ =0, the effective factor $\bar{m}/(x_v m_v)$ for gravity has changed due to the mole fraction, and mean molecular weight.

In the stratosphere where $T$ and $x_n$ are almost constant,  the integration of  Eq. (5) becomes  
\begin{equation}
  \tau =(\kappa_vx_vm_v+\kappa_nx_nm_n)\frac{p}{\bar{m}g}.
\end{equation}
The stratosphere is in a radiative equilibrium and the temperature structure is given as 
\begin{equation}
  \pi B =\frac{1}{2}F_{IRtop}\left(1+\frac{3}{2}\tau \right),
\end{equation} 
where $B=\sigma T^4/\pi$ is the blackbody radiation intensity.
At the tropopause, we have
\begin{equation}
  \frac{1}{2}F_{IRtop}\left(1+\frac{3}{2}\tau_{tp} \right)=\sigma T_{tp}^4,
\end{equation} 
and
\begin{equation}
  \tau_{tp}=\kappa_vp^{\ast}(T_{tp})\frac{1}{g}\frac{m_v}{\bar{m}}.
\end{equation} 

There is a maximum value of $F_{IRtop} (=F_{KI-limit})$ which will satisfy the above two equations which is dnoted Komabayashi-Ingersoll limit (Nakajima et al. 1992).



\section{Applied to H$_2$ background atmosphere}

It is investigated the gray calculations for cases of background gases with H$_2$ for $g=9.8 $m/s$^2$.  The relationships between $T_s$ and $F_{IRtop}$ for several initial $p_{H2\ 0}$ cases are presented in Fig. 1.
\begin{figure}[htbp]
\begin{center}
\includegraphics[keepaspectratio,scale=0.4]{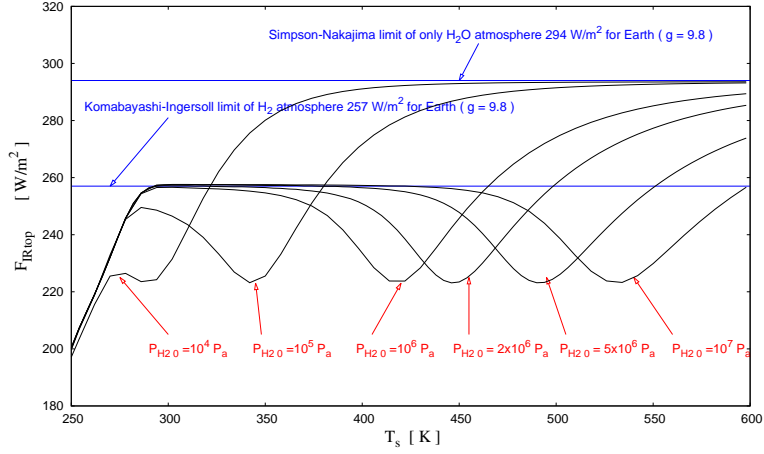}
\vspace*{0.5cm}
\caption{The relationships between $T_s$ and $F_{IRtop}$ 
for several initial $p_{H2\ 0} (=10^4, 10^5, 10^6, 2\times10^6, 5\times10^6, $ and $10^7$ Pa)  cases of H$_2$. }
\end{center}
\end{figure}

For H$_2$ background atmosphere, it is characteristic that $F_{IRtop}$ increases with surface temperature $T_s$ at the first time, however, it seems to reach the upper limit, then decrease down and again increase to reach the second saturated limit denoted as the SN-limit (Goldblatt \& Watson 2012). 

 The first upper limit is due to the dominant H$_2$ background atmosphere, where the optical depth is unity at the gas temperature $T \simeq 260 \sim 280$ K as shown in  Figs. 2 and 3 where the case of $P_{H2\ 0}=10^6$ Pa is presented.  The decrease of $F_{IRtop}$ after the first upper limit is due to the temperature decrease of the gas temperature corresponding to the region where the optical depth is about unity (later we call it as unity optical depth).  After the lower limit of  $F_{IRtop}$, it increases due to the temperature increase of the gas temperature corresponding to the unity optical depth which is shown in Fig. 2 and magnified in Fig. 3, where the detailed feature could be seen.  The $F_{IRtop}$ then increases to the SN-limit as shown in Fig. 1.

\subsection {Moist tropospheric asymptotic limit (SN-limit)}
  There is a limit of $F_{IRtop}$ that the initial $p_0$ of the background atmosphere is small or null where the atmosphere is consisted almost only of saturated H$_2$O vapour.  This is called the "Moist tropospheric asymptotic limit", "Saturated vapour limit",  "SN-limit" (Goldblatt \& Watson  2012), or "steam limit" (Koll \& Cronin 2019).  
  The second upper limit is the same with the only water vapour atmosphere, being the case of saturated vapour pressure ($\simeq $ 294 W/m$^2$ for $g=9.8$m/s$^2$ and $\simeq$ 340 W/m$^2$ for super-Earth in Fig. 10).

 When the surface temperature is enough high, the mole fraction of water vapour is almost unity and optical depth increases.  When the optical depth is smaller than unity, the outgoing flux comes from the bottom of the Hydrogen atmosphere.  However, when optical depth is greater than unity, the outgoing flux comes from the place where the optical depth is almost unity from the top of the atmosphere ( UOD: unity optical depth).  Then the outgoing flux is almost the same as that of a pure H$_2$O atmosphere.

\begin{figure}[htbp]
\begin{center}
\includegraphics[keepaspectratio,scale=0.4]{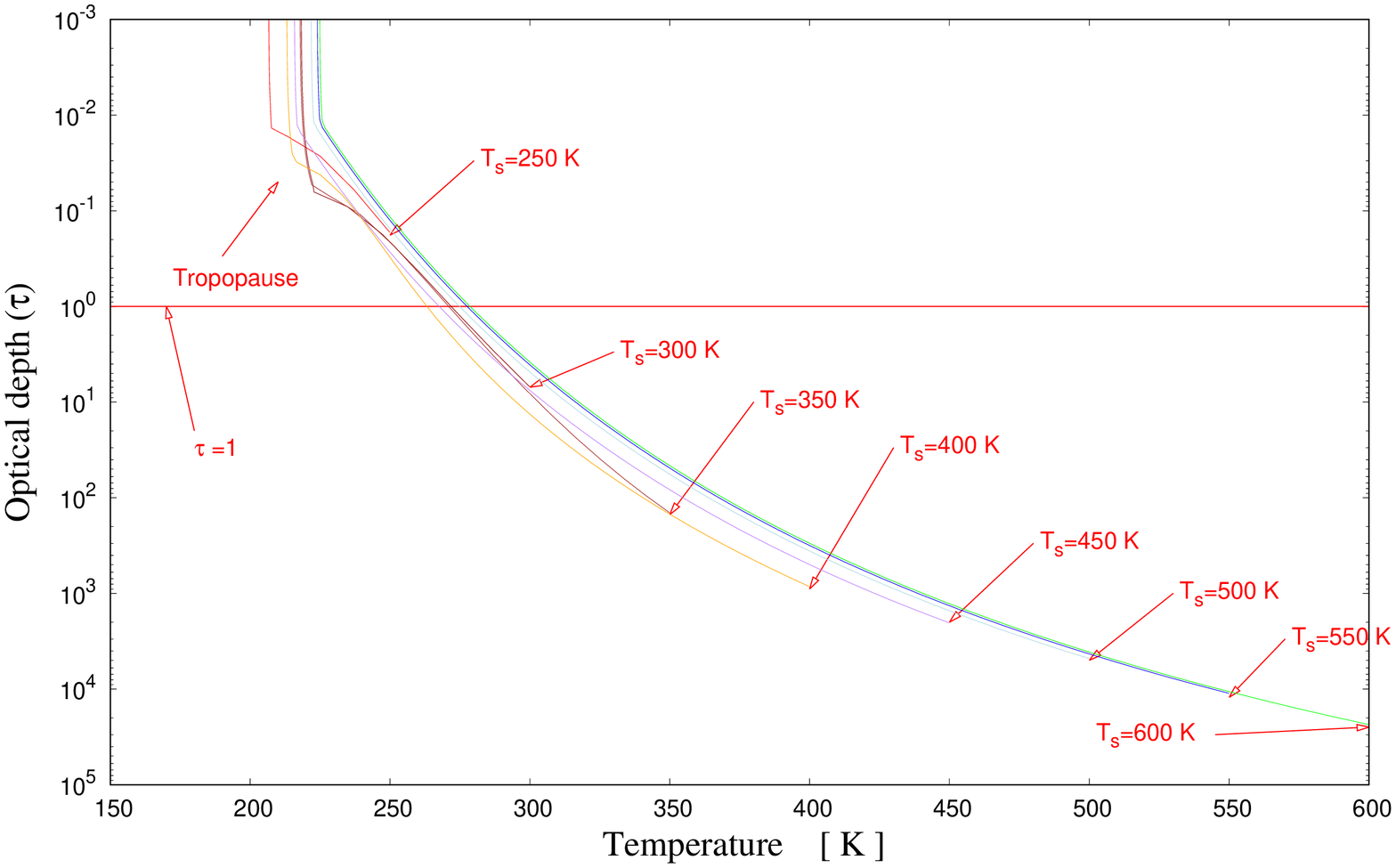}
\vspace{0.5cm}
\caption{The relationships between $T_s$ and optical depth $(\tau)$ of $p_{H2 \ 0}=10^6$ Pa for several $T_s$ cases.}
\end{center}
\end{figure}

The relationships between surface temperature $T_s$ and optical depth $(\tau )$ of $p_{H2\ 0}=10^6$ Pa for several $T_s$ cases are presented in Fig. 2.  The cases of $T_s =250$K, 300K, 350K, 400K, 450K, 500K, 550K, and 600K are presented by red, dark-red, brown, orange, purple, light-blue, blue, and green colour curve, respectively, The unit optical depth ($\tau=1$) is also presented by the red line and the magnified figure is presented in Fig. 3.

\begin{figure}[htbp]
\begin{center}
\includegraphics[keepaspectratio,scale=0.4]{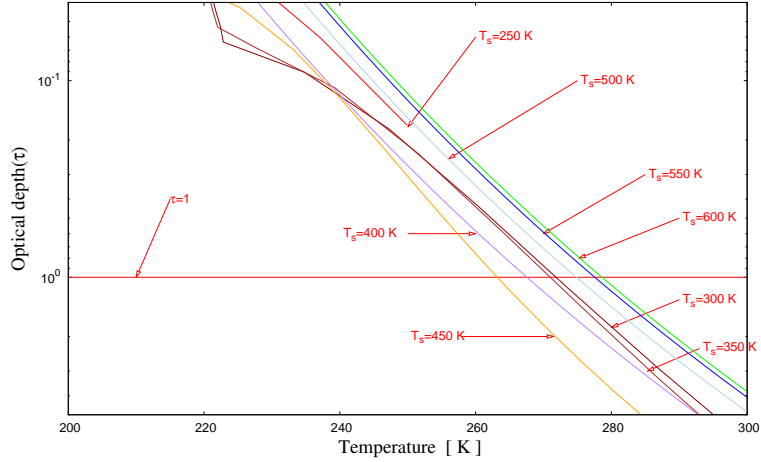}
\vspace*{0.5cm}
\caption{The relationships between $T_s$ and optical depth $(\tau )$ of $p_{H2\ 0}=10^6$ Pa for several $T_s$ cases are presented.  It is the magnified figure of Fig. 2 to explain the change of $F_{IRtop}$ for different of $T_s$. }
\end{center}
\end{figure}

\begin{figure}[htbp]
\begin{center}
\includegraphics[keepaspectratio,scale=0.4]{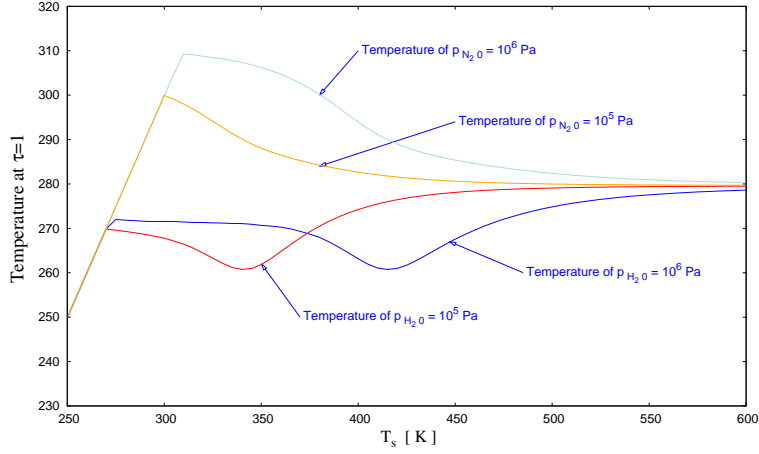}
\vspace*{0.5cm}
\caption{The relationships between the temperature of UOD (Unity optical depth) and $T_s$ of $p_{H_2 \ 0}= 10^6$ and $10^5 $ Pa  are presented by blue and red curves, respectively.  The cases for N$_2$ atmospheres of $p_{N_2 \ 0}= 10^6$ and $10^5 $ Pa  are also presented for reference by light blue and orange curves, respectively. }
\end{center}
\end{figure}

The luminosity is mainly radiated at around $\tau \simeq 1$ and the intensity is related to the local temperature of $\tau \simeq 1$.  The first bump of $p_{H2 \ 0}=10^6$ in Fig. 1  is appeared at $T_s=300 \sim 350 $K and then the luminosity decreases around  $T_s=400 \sim 450 $K  and then increases around $T_s=500 \sim 600 $K.  The detailed features could be seen in Fig. 3 as the line of $\tau=1$ is crossed by each $T_s$ curve.

 The features could be understood in Fig. 4, where the relationship between the temperature at UOD (Unity optical depth) and $T_s$ of $p_{H_2 \ 0}= 10^6 $ Pa  is presented by blue curves.  The luminosity decreases around  $T_s=400 \sim 450 $K and then increases around $T_s=500 \sim 600 $K.  The case of $p_{H_2 \ 0}= 10^5$ Pa is shown there by red curve.  The cases for N$_2$ atmospheres of $p_{N_2 \ 0}= 10^6$ and $10^5$ Pa  are also presented for reference by light blue and orange curves, respectively.  For the case that the total optical depth is smaller than unity, the temperature of UOD is assumed to be equal to $T_s$.

\begin{figure}[htbp]
\begin{center}
\includegraphics[keepaspectratio,scale=0.4]{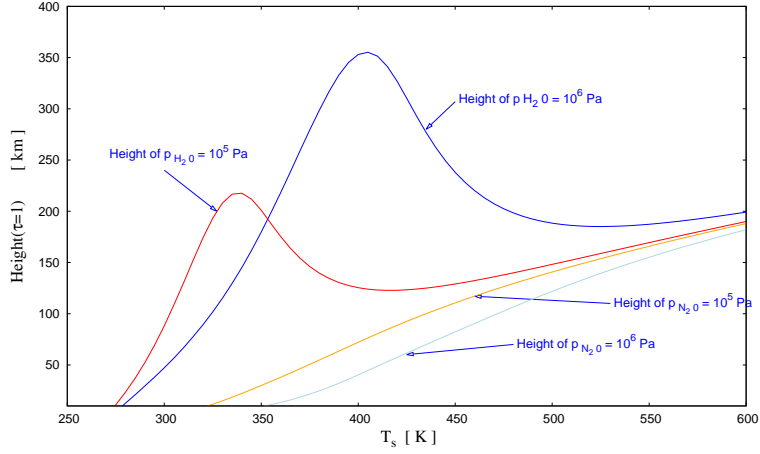}
\vspace*{0.5cm}
\caption{The relationships between the height of UOD and $T_s$ of $p_{H_2 \ 0} = 10^6$ and $10^5$ Pa  are presented by blue and red curves, respectively.  The cases for N$_2$ atmospheres of $p_{N_2 \ 0}= 10^6$ and $10^5$ Pa  are also presented for reference by light blue and orange curves, respectively.  The height is zero for the case that the total optical depth is smaller than unity.  }
\end{center}
\end{figure}

In Fig. 5, the relationships between the height of UOD and $T_s$ of $p_{H_2 \ 0} = 10^6$ is presented by blue and curve. The height of UOD has increased to $\sim 350$ km, where the temperature of UOD has decreased to $\sim$ 260 K (see Fig. 4).  The case of $p_{H_2 \ 0}= 10^5$ Pa is shown there by red curve.  The cases for N$_2$ atmospheres of $p_{N_2 \ 0}= 10^6$ and $10^5$ Pa  are also presented for reference by light blue and orange curves, respectively.  The height of UOD is taken to be zero for the case that the total optical depth is smaller than unity.

In the low temperature limit, the lapse rate is assumed to be the dry adiabat, $dT/dz=-g/c_p$, as described in Pierrehumbert (2010) and Appendix B
\begin{equation} 
  H_T^{dilute}=\frac{c_p T}{g} \propto 1/M,
\end{equation} 
where $c_p$ is the specific heat per unit mass and proportional to the inverse of the molecular weight $M$.  In Appendix B, it is investigated the analytical expression derived by Koll \& Cronin (2019) for the OLR limit values of the steam limit and dilute limit.  In Appendix C, it is tried to derive Eq. (33) in Koll \& Cronin (2019) for suspected typos.

It should be noted that at $T_s \sim$ 400 K for $p_{H_2 \ 0}= 10^6$ Pa the height of UOD has increased to $H_T^{UOD} \simeq 350$ km in H$_2$, whereas $H_T^{UOD} \simeq 40$ km in N$_2$. The main difference must be due to the low molecular weight of H$_2$ relative to N$_2$. 

The increase of the UOD height is related to the features that $F_{IRtop}$ increases with surface temperature $T_s$ at first time then decreases down and again increase to reach the second saturated limit.  Thr first $F_{IRtop}$ increase is due to the increase of $T_s$ and the lower value of  the water vapoure (Figs. 6 and 7).  The $F_{IRtop}$ upper limit is due to the KI-limit.  The  $F_{IRtop}$ decrease is due to the increase of the water vapour and the optical depth (Figs. 6, 7, and 8).  The UOD height increase is related to the decrease the UOD temperature (Figs. 4 and 5).  The second saturated limit is due to the increase and saturation limit of the water vapoure (Figs. 4 $\sim $8).

\begin{figure}[htbp]
\begin{center}
\includegraphics[keepaspectratio,scale=0.4]{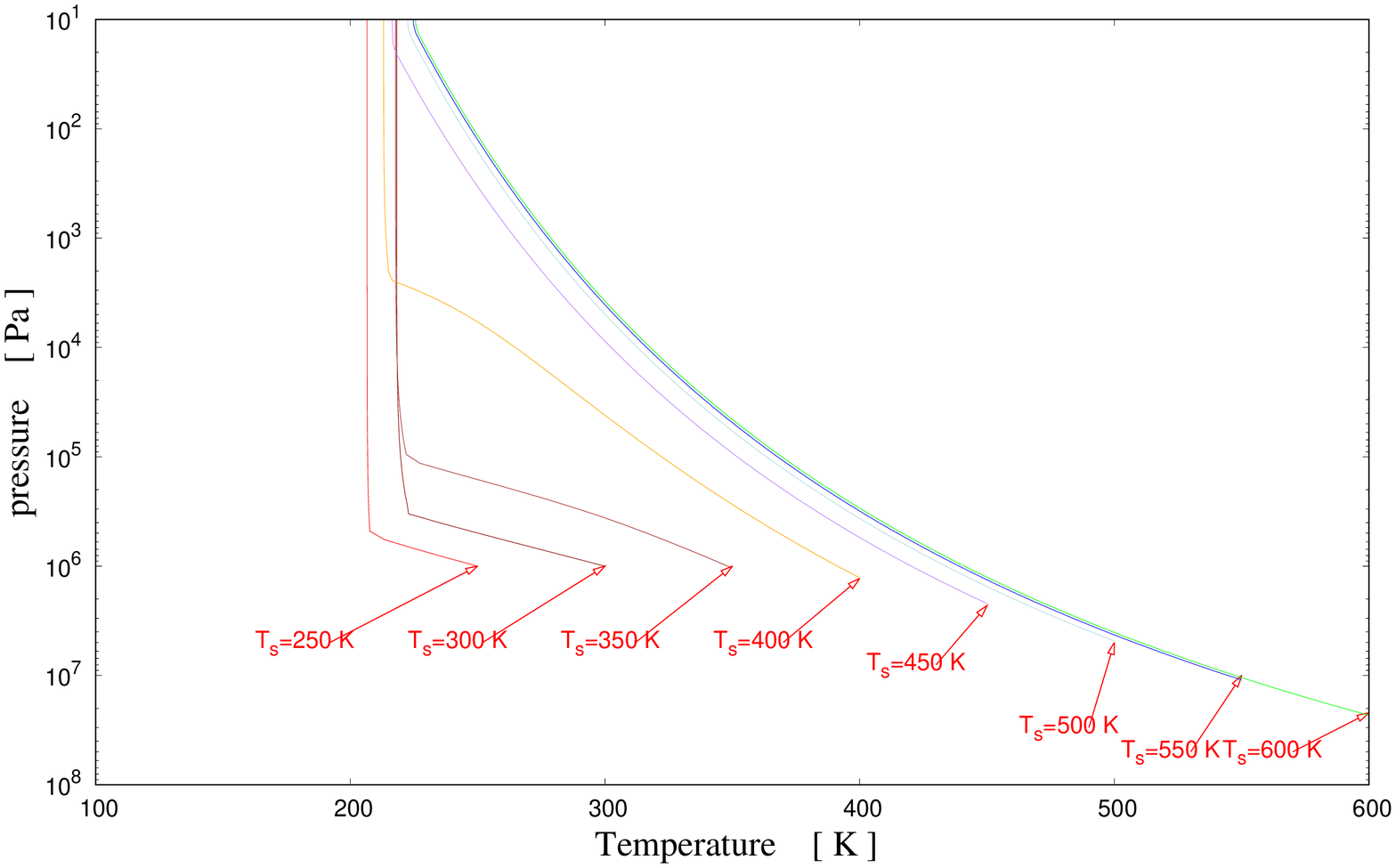}
\vspace*{0.5cm}
\caption{The relationships between temperature and pressure for several $T_s$ cases of $p_{H_2\ 0}=10^6$ Pa. }
\end{center}
\end{figure}

The relationships between temperature and pressure of $p_{H2\ 0}=10^6$ Pa for several $T_s$ cases are calculated in Fig. 6 and 
the relationships between mole fraction and pressure are shown in Fig. 7.  In Fig. 8, the relationships between mole fraction and optical depth are presented. 
\begin{figure}[htbp]
\begin{center}
\includegraphics[keepaspectratio,scale=0.4]{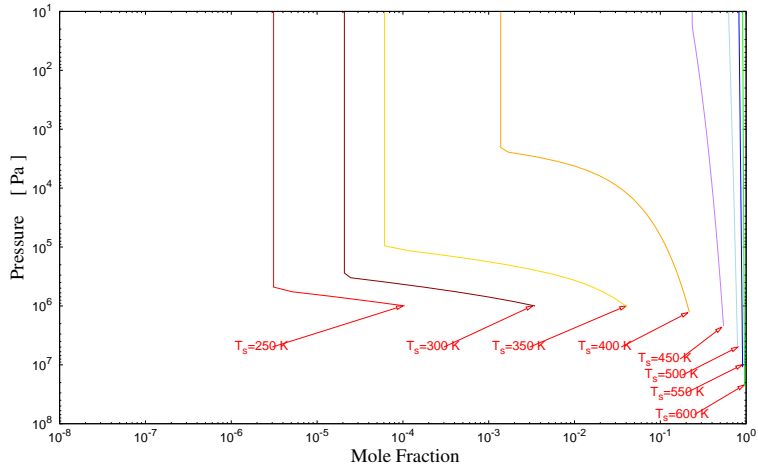}
\vspace*{0.5cm}
\caption{The relationships between the mole fraction of the water vapour and the pressure of $p_{H_2\ 0}=10^6$ Pa for several $T_s$ cases. }
\end{center}
\end{figure}

\begin{figure}[htbp]
\begin{center}
\includegraphics[keepaspectratio,scale=0.4]{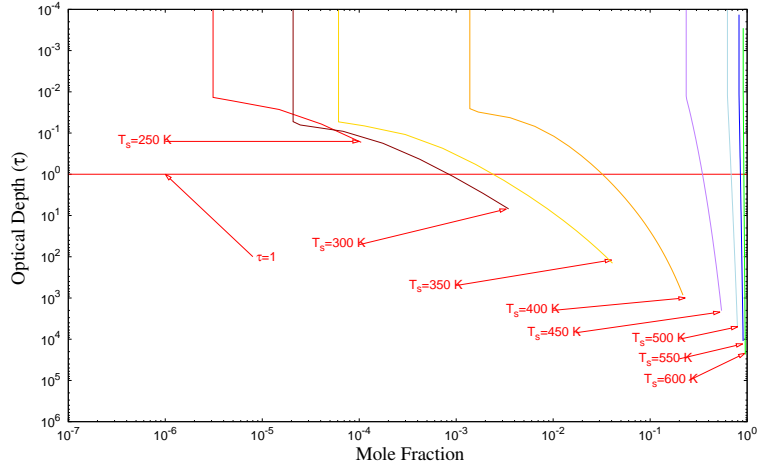}
\vspace*{0.5cm}
\caption{The relationships between the mole fraction and the optical depth  for several $T_s$ cases of $p_{H_2\ 0}=10^6$ Pa. }
\end{center}
\end{figure}

For H$_2$ atmosphere, optical depth $\tau$ at tropopause is given by Eq. (9) where 
the factor $m_v/\bar{m}$ increases about 10 from N$_2$ to H$_2$  atmosphere which causes the decrease of KI-limit ($\simeq$ 420 W/m$^2$  (N$_2$) to $\simeq$ 257 W/m$^2$ (H$_2$), as shown in Fig. 9).

The difference between N$_2$ and H$_2$ background atmosphere is the difference of the mean molecular weight which corresponds to the change of the optical depth.  For H$_2$ atmosphere, the optical depth has increased.  Even for the same surface temperature, the temperature at optical depth $\tau=1$ decreases for H$_2$ atmosphere.  Then the outgoing radiation has decreased for the H$_2$ atmosphere compared to the N$_2$ atmosphere. 

If H$_2$ pressure increases, the outline of the graph has moved to the right-hand which is shown in Fig. 1.  If the surface temperature increased to the right-hand direction, there is a possibility that the ocean would evaporate.  

The relationships between $T_s$ and $F_{IRtop}$ 
for several $p_{H_2\ 0}$ and $p_{N_2\ 0}$ cases are presented in Fig. 9.  Some H$_2$ results in Fig. 1 are included in Fig. 9. 
The cases of N$_2$ are almost the same given in Nakajima et al. 1992, except that the molecular weight of  N$_2$ is taken 28 (it is taken 18 in Nakajima et al. (1992) for its simplicity), then KI-limit has increased from $\simeq$ 385 W/m$^2$ to $\simeq$ 420 W/m$^2$.

The cases of H$_2$ for super-Earth ($g=9.8\times 2 $ \ m/s$^2$) are shown in Fig. 10.
\begin{figure}[htbp]
\begin{center}
\includegraphics[keepaspectratio,scale=0.4]{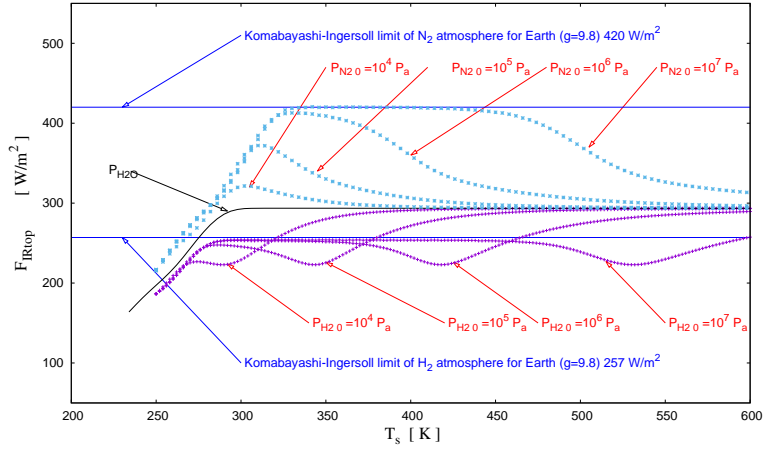}
\vspace*{0.5cm}
\caption{It is compared the gray calculations for cases of background gases with H$_2$ and N$_2$ for $g=9.8$m/s$^2$.  The relationships between $T_s$ and $F_{IRtop}$ 
for several initial $p_{H2 \ 0}$ and $p_{N2\ 0}$ cases are presented. Some H$_2$ results in Fig. 1 are included.  The arrow of $ P_{H2O}$ shows the pure H$_2$O case and SN-limit ($\simeq$ 295 W/m$^2$).}
\end{center}
\end{figure}

\begin{figure}[htbp]
\begin{center}
\includegraphics[keepaspectratio,scale=0.4]{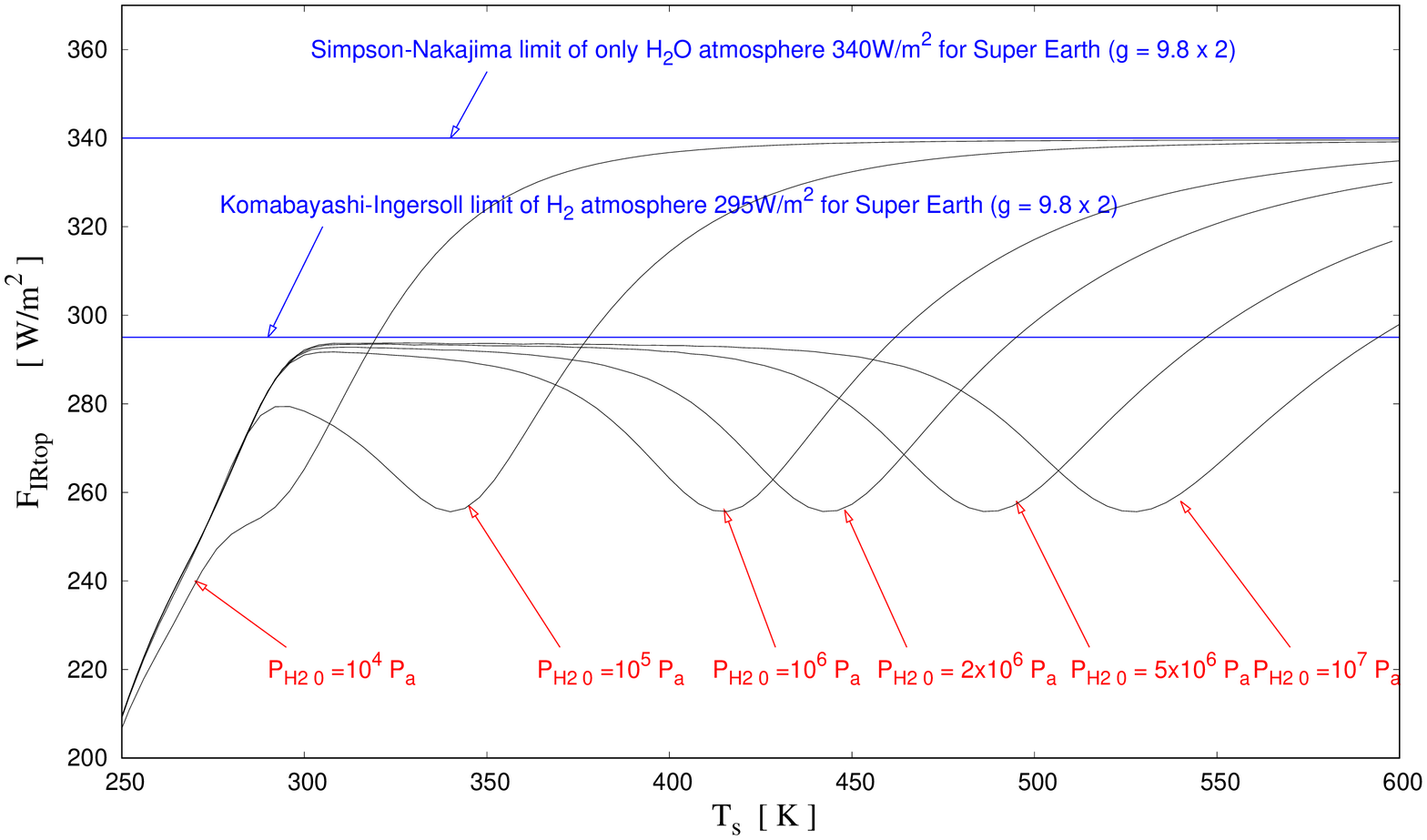}
\vspace*{0.5cm}
\caption{For super-Earth cases with H$_2$ atmosphere, the relationships between $T_s$ and $F_{IRtop}$ 
for several initial $p_{H2\ 0}$ cases are presented. }
\end{center}
\end{figure}

\begin{figure}[htbp]
\begin{center}
\includegraphics[keepaspectratio,scale=0.4]{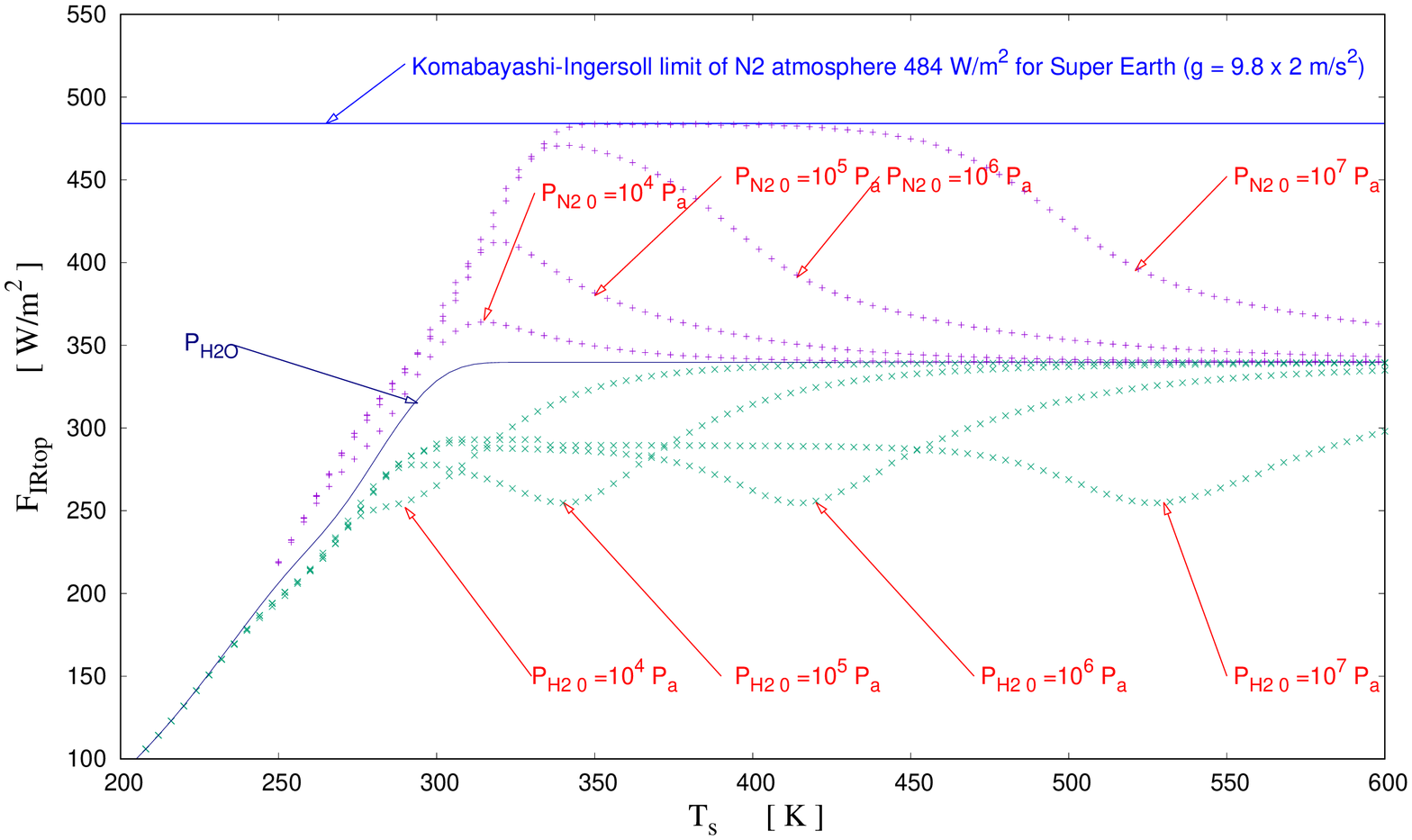}
\vspace*{0.5cm}
\caption{For super-Earth cases of background gases with H$_2$ and N$_2$, the relationships between $T_s$ and $F_{IRtop}$ in several initial $p_{H2\ 0}$ cases are presented.  The arrow of $P_{H2O}$ shows the case of $p_{H2\ 0}=0$ Pa which means only H$_2$O atmosphere component corresponding to SN-limit.}
\end{center}
\end{figure}

In Fig. 11, super-Earth cases of background gases with H$_2$ and N$_2$ are presented for the relationships between $T_s$ and $F_{IRtop}$ in several initial $p_{H2\ 0}$ cases. The arrow of $P_{H2O}$ shows the case of $p_{H2\ 0}=0$ Pa which means only H$_2$O atmosphere component corresponding to the SN-limit.


\section{Applied to ${\rm He}$ background atmosphere}
The situations of He background atmosphere  are similar to the case of H$_2$ background atmosphere except the treatment of the mean molecular weight and the specific heat of He as $c_p=2.5$ R, being mono atomic gas, compared to that of H$_2$, as $c_p=3.5$ R, being dipole atomic gas. 
\begin{figure}[htbp]
\begin{center}
\includegraphics[keepaspectratio,scale=0.4]{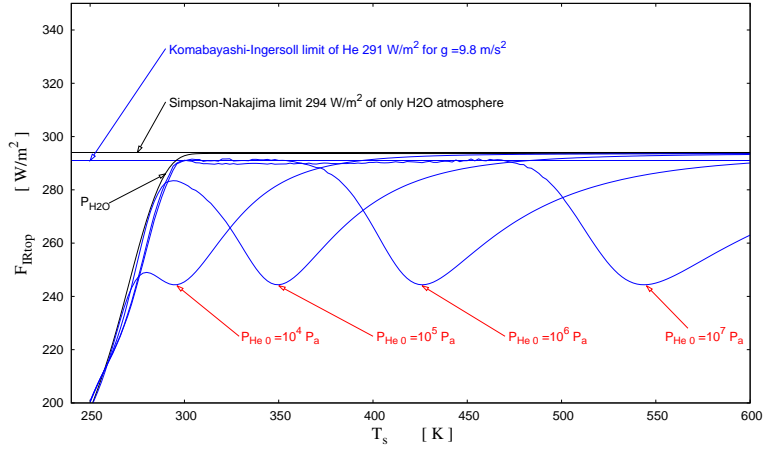}
\vspace*{0.5cm}
\caption{For He background gases of $g=9.8$m/s$^2$, the  relationships between $T_s$ and $F_{IRtop}$ 
for several initial $p_{He\ 0}$ cases are presented.  The value of KI-limit for He atmosphere is $\simeq$ 291 W/m$^2$ which is almost the same value of the SN-limit $\simeq$ 294 W/m$^2$.}
\end{center}
\end{figure}

In Fig. 12, the results of He atmosphere are similar to those of H$_2$ cases in Fig. 1.  The first upper limit of $p_{He\ 0}=10^6$ and $10^7$ are the KI-limit of He atmosphere.  Then they decrease down and increase approximately to the SN-limit as the cases of H$_2$ background gas as presented in Fig. 1.

It must be noticed that the KI-limit for He atmosphere is $\simeq$ 291 W/m$^2$ which is almost the same value of the SN-limit $\simeq$ 294 W/m$^2$.  It seems to be a  problem that the SN-limit is greater than the KI-limit, however it is not a problem because the components of the background atmospheres have changed from He to H$_2$O.  It is almost the same for H$_2$ atmosphere cases in Fig. 1.

It is pointed out by Koll \& Cronin (2019) that H$_2$ is the only background gas for which the dilute runaway (KI-limit) lies below the steam limit (SN-limit).  However as shown in Fig 12, He is also the background gas for which the dilute limit lies below the steam limit.
 
\begin{figure}[htbp]
\begin{center}
\includegraphics[keepaspectratio,scale=0.4]{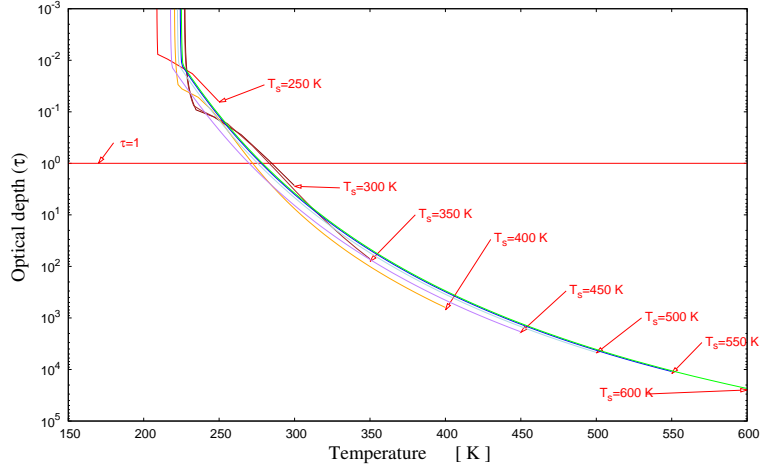}
\vspace*{0.5cm}
\caption{The relationships between $T_s$ and optical depth $(\tau)$ for several $T_s$ cases of $p_{He\ 0}=10^6$ Pa are presented. }
\end{center}
\end{figure}

The relationships between $T_s$ and optical depth $(\tau)$ of $p_{He\ 0}=10^6$ Pa for several $T_s$ cases are presented in Fig. 13.  The $T_s$ =250K, 300K, 350K, 400K, 450K, 500K, 550K, and 600K are presented by red, dark-red, brown, orange, purple, light-blue, blue, and green colour curves, respectively, The unit optical depth ($\tau=1$) is also presented in red line.  

\begin{figure}[htbp]
\begin{center}
\includegraphics[keepaspectratio,scale=0.4]{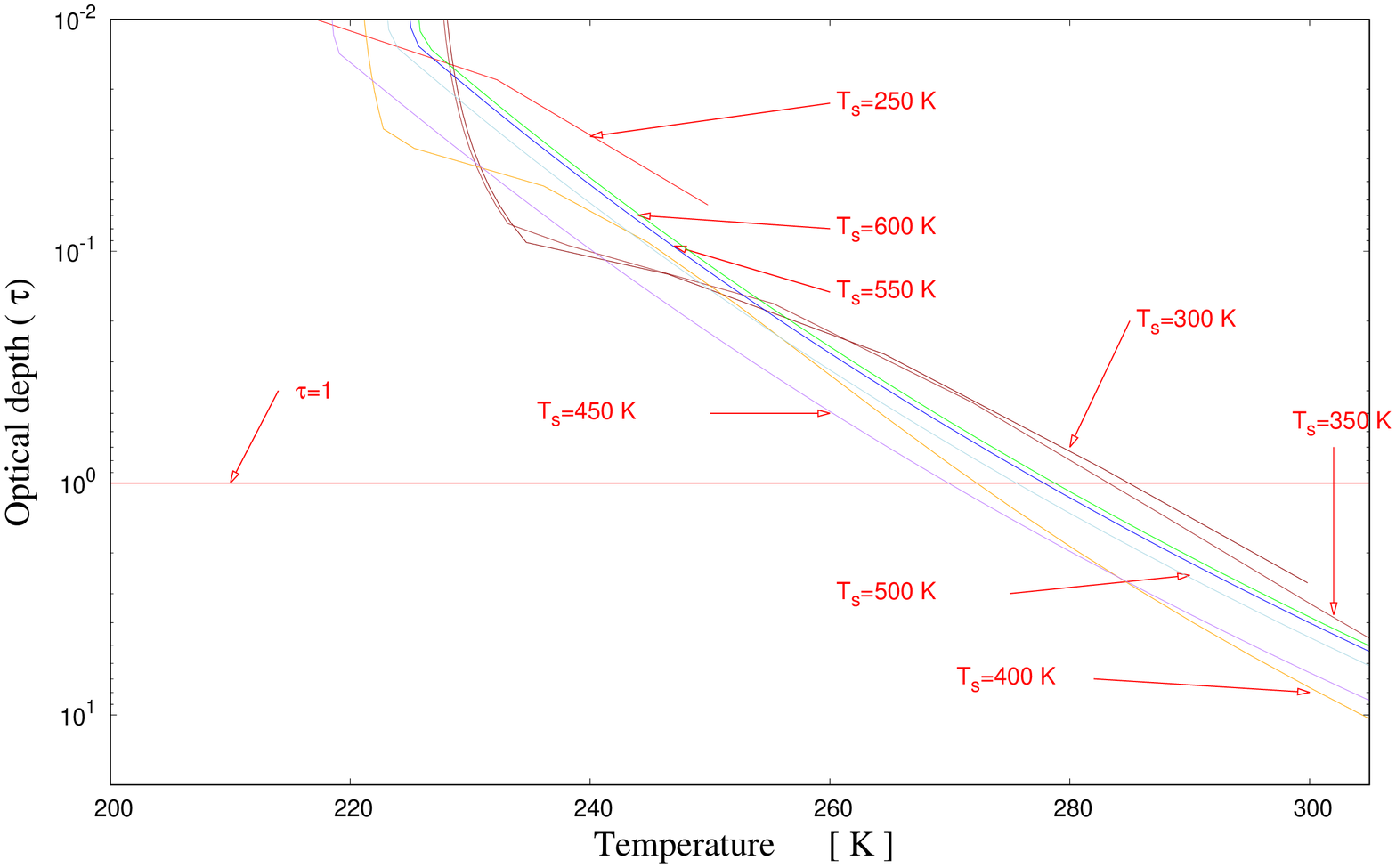}
\vspace*{0.5cm}
\caption{The relationships between $T_s$ and optical depth $(\tau )$ of $p_{He\ 0}=10^6$ Pa for several $T_s$ cases.  It is the magnified figure of Fig. 13.}
\end{center}
\end{figure}

Fig. 14 is the magnified figure of Fig. 13 to explain the change of $F_{IRtop}$ for different of $T_s$.  
The luminosity is mainly radiated at around $\tau \simeq1$ and the intensity is related to the local temperature of $\tau \simeq 1$.  The first bump of $P_{He\ 0}=10^6$ Pa in Fig. 12  is appeared at $T_s \simeq 300 \sim 350$ K and then the luminosity decreases around  $T_s \simeq 400 \sim 450$ K and then increases around $T_s \simeq 500 \sim 600$ K.  

\begin{figure}[htbp]
\begin{center}
\includegraphics[keepaspectratio,scale=0.4]{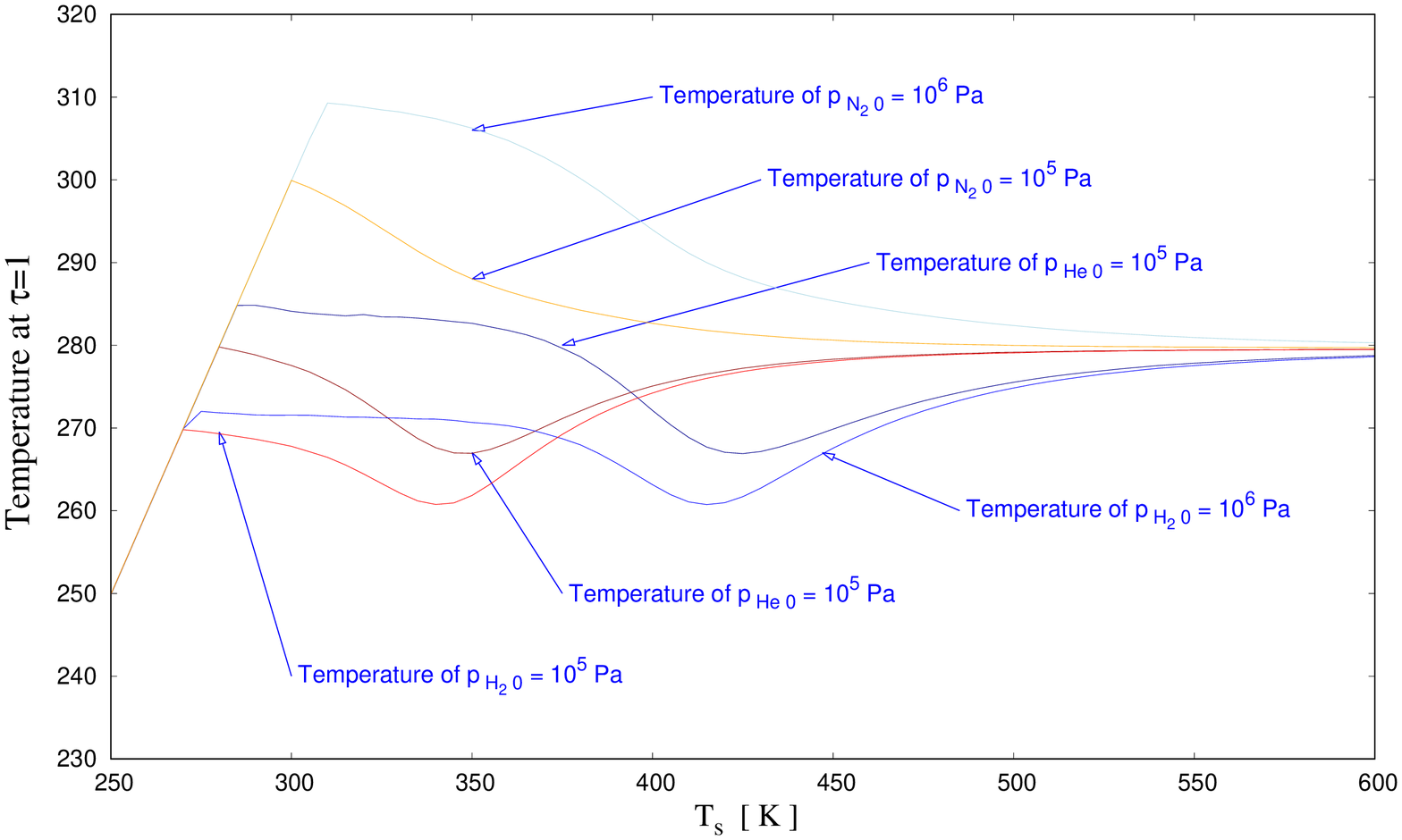}
\vspace*{0.5cm}
\caption{The relationships between the temperature of UOD (Unity optical depth) and $T_s$ of $p_{He \ 0}= 10^6$ and $10^5 $ Pa.}
\end{center}
\end{figure}

The features could be understood in Fig. 15, where the relationship between the temperature at UOD (Unity optical depth) and $T_s$ of $p_{He \ 0}= 10^6 $ Pa  is presented by dark blue curves.  The luminosity decreases around  $T_s=400 \sim 450 $K and then increases around $T_s=500 \sim 600 $K.  The case of $p_{He \ 0}= 10^5$ Pa is shown there by dark red curve.  The cases for H$_2$ atmospheres of $p_{H_2 \ 0}= 10^6$ and $10^5$ Pa  are also presented by blue and red curves, respectively.  The cases for N$_2$ atmospheres of $p_{N_2 \ 0}= 10^6$ and $10^5$ Pa  are also presented for reference by light blue and orange curves, respectively.  For the case that the total optical depth is smaller than unity, the temperature of UOD is assumed to be equal to $T_s$.

In Fig. 16, the relationships between the height of UOD and $T_s$ of $p_{He \ 0} = 10^6$ is presented by dark blue curve. The height of UOD has increased to $\sim 250$ km, where the temperature of UOD has decreased to $\sim$ 270 K (see Fig. 15).  The case of $p_{He \ 0}= 10^5$ Pa is shown there by dark red curve.  The cases for H$_2$ atmospheres of $p_{N_2 \ 0}= 10^6$ and $10^5$ Pa  are also presented by blue and red curves, respectively.  The cases for N$_2$ atmospheres of $p_{N_2 \ 0}= 10^6$ and $10^5$ Pa  are also presented for reference by light blue and orange curves, respectively.  The height of UOD is taken to be zero for the case that the total optical depth is smaller than unity.

The relationships between temperature and pressure of $p_{He\ 0}=10^6$ Pa for several $T_s$ cases are presented in Fig. 17, and the relationships between mole fraction and pressure are shown in Fig. 18  The relationships between mole fraction and optical depth are calculated in Fig. 19.

\begin{figure}[htbp]
\begin{center}
\includegraphics[keepaspectratio,scale=0.4]{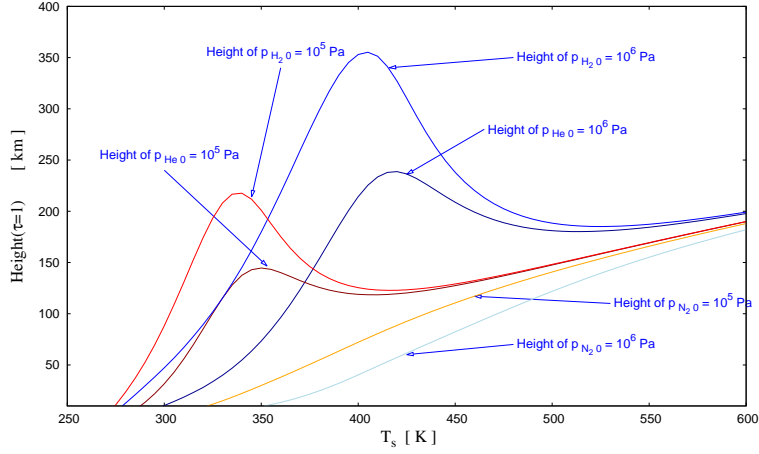}
\vspace*{0.5cm}
\caption{The relationships between the height of UOD and $T_s$ of $p_{He \ 0} = 10^6$ and $10^5$ Pa  are presented by dark blue and dark red curves, respectively. 
 The cases for H$_2$ atmospheres of $p_{H_2 \ 0}= 10^6$ and $10^5$ Pa  and the cases for N$_2$ atmospheres of $p_{N_2 \ 0}= 10^6$ and $10^5$ Pa  are also presented for reference.}
\end{center}
\end{figure}

\begin{figure}[htbp]
\begin{center}
\includegraphics[keepaspectratio,scale=0.4]{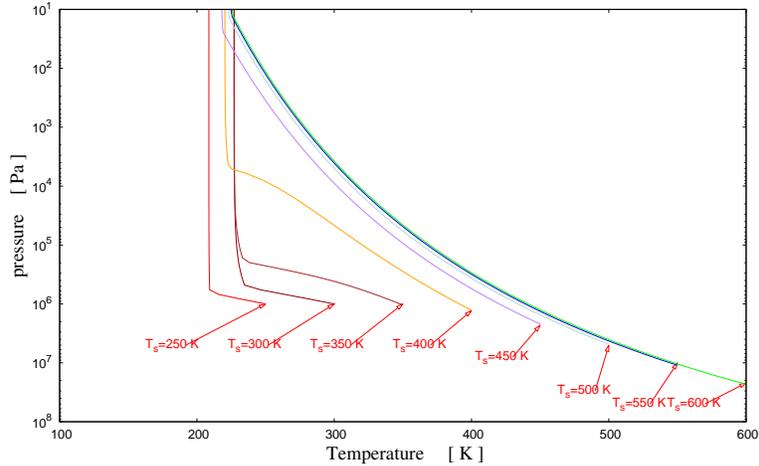}
\vspace*{0.5cm}
\caption{The relationships between temperature and pressure of $p_{He\ 0}=10^6$ Pa for several $T_s$ cases with He. }
\end{center}
\end{figure}
 
\begin{figure}[htbp]
\begin{center}
\includegraphics[keepaspectratio,scale=0.4]{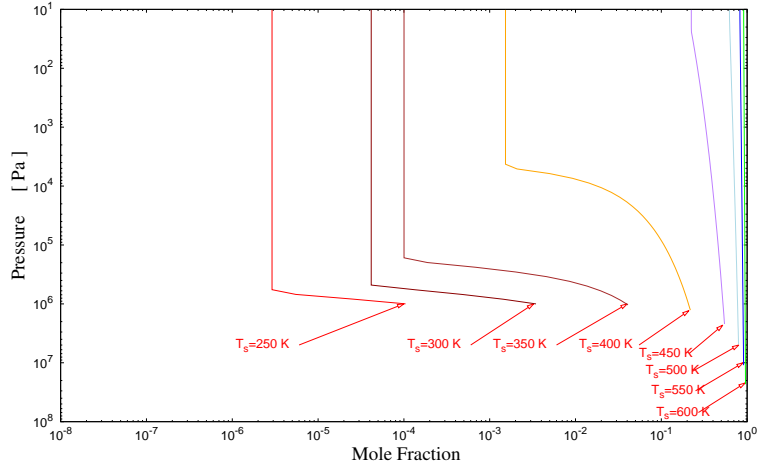}
\vspace*{0.5cm}
\caption{The relationships between mole fraction and pressure of $p_{He\ 0}=10^6$ Pa for several $T_s$ cases with He}
\end{center}
\end{figure}

\begin{figure}[htbp]
\begin{center}
\includegraphics[keepaspectratio,scale=0.4]{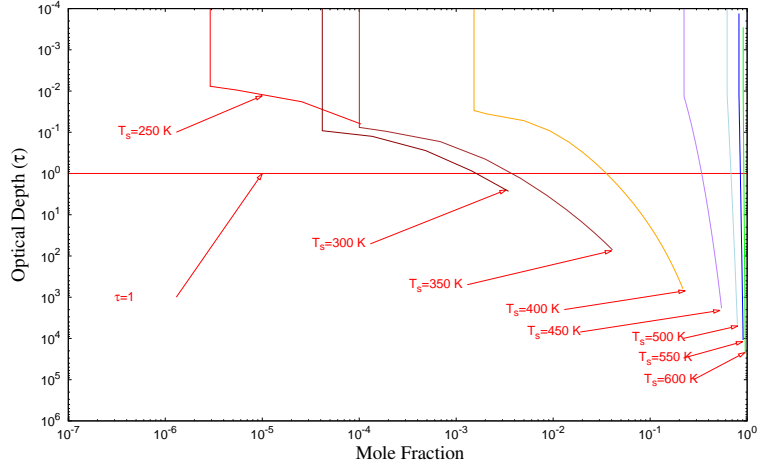}
\vspace*{0.5cm}
\caption{For He in $g=9.8m/s^2$, the relationships between mole fraction and optical depth of $p_{He 0}=10^6$ Pa for several $T_s$ cases are presented. }
\end{center}
\end{figure}

\begin{figure}[htbp]
\begin{center}
\includegraphics[keepaspectratio,scale=0.4]{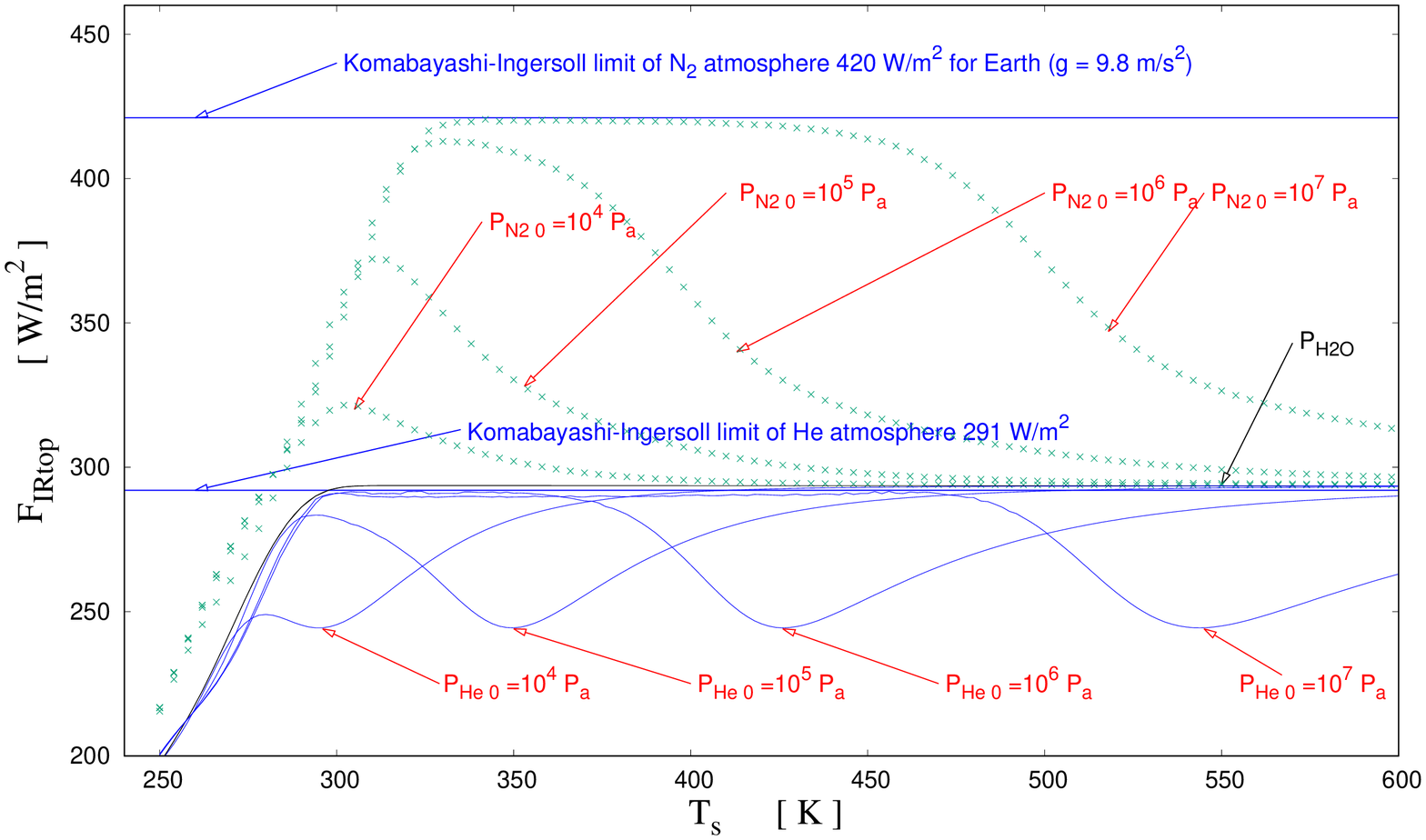}
\vspace*{0.5cm}
\caption{The relationships between $T_s$ and $F_{IRtop}$ 
for several $p_{He\ 0}$ and $p_{N2\ 0}$ cases of background gases with He and N$_2$.}
\end{center}
\end{figure}

To compare the cases of N$_2$, the results of N$_2$ and He are presented in Fig. 20 for $g$=9.8m/s$^2.$  
The relationships between $T_s$ and $F_{IRtop}$ 
for several $p_{He\ 0}$ and $p_{N2\ 0}$ cases are presented there.  
Several cases of $p_{He\ 0}$ in Fig. 12 are also included here.   
The case of pure H$_2$O atmosphere is shown by black curve. 

For super-Earth in $g=9.8\times 2$ m/s$^2$, it is investigated the gray calculations for cases of background gases with He in Fig 21.  The relationships between $T_s$ and $F_{IRtop}$ for several $p_{He\ 0}$ cases are presented.  The case of pure H$_2$O atmosphere is shown by black curve.  It is noticed that KI and SN-limits are different from Earth cases in Fig. 12.

It is investigated the gray calculations for cases of background gases with N$_2$ and He for super-Earth case in Fig. 22.  The relationships between $T_s$ and $F_{IRtop}$ for several $p_{N2\ 0}$ and $p_{He\ 0}$ cases are presented there.  Several cases of $p_{He\ 0}$ in Fig. 21 are also included here.

\begin{figure}[htbp]
\begin{center}
\includegraphics[keepaspectratio,scale=0.4]{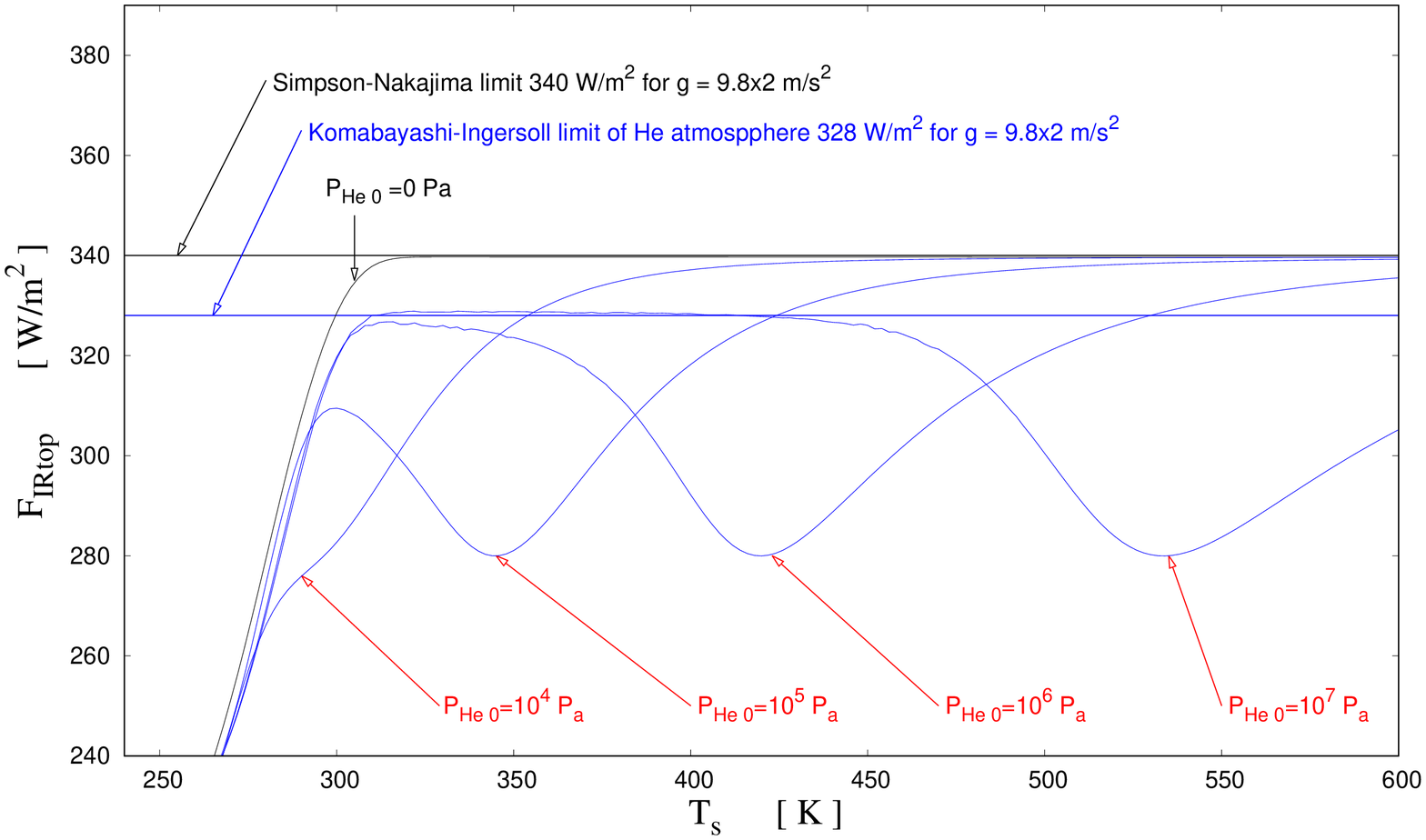}
\vspace*{0.5cm}
\caption{For super-Earth in $g=9.8\times 2$ m/s$^2$, the relationships between $T_s$ and $F_{IRtop}$ for several $p_{He\ 0}$ cases are presented.}
\end{center}
\end{figure}

\begin{figure}[htbp]
\begin{center}
\includegraphics[keepaspectratio,scale=0.4]{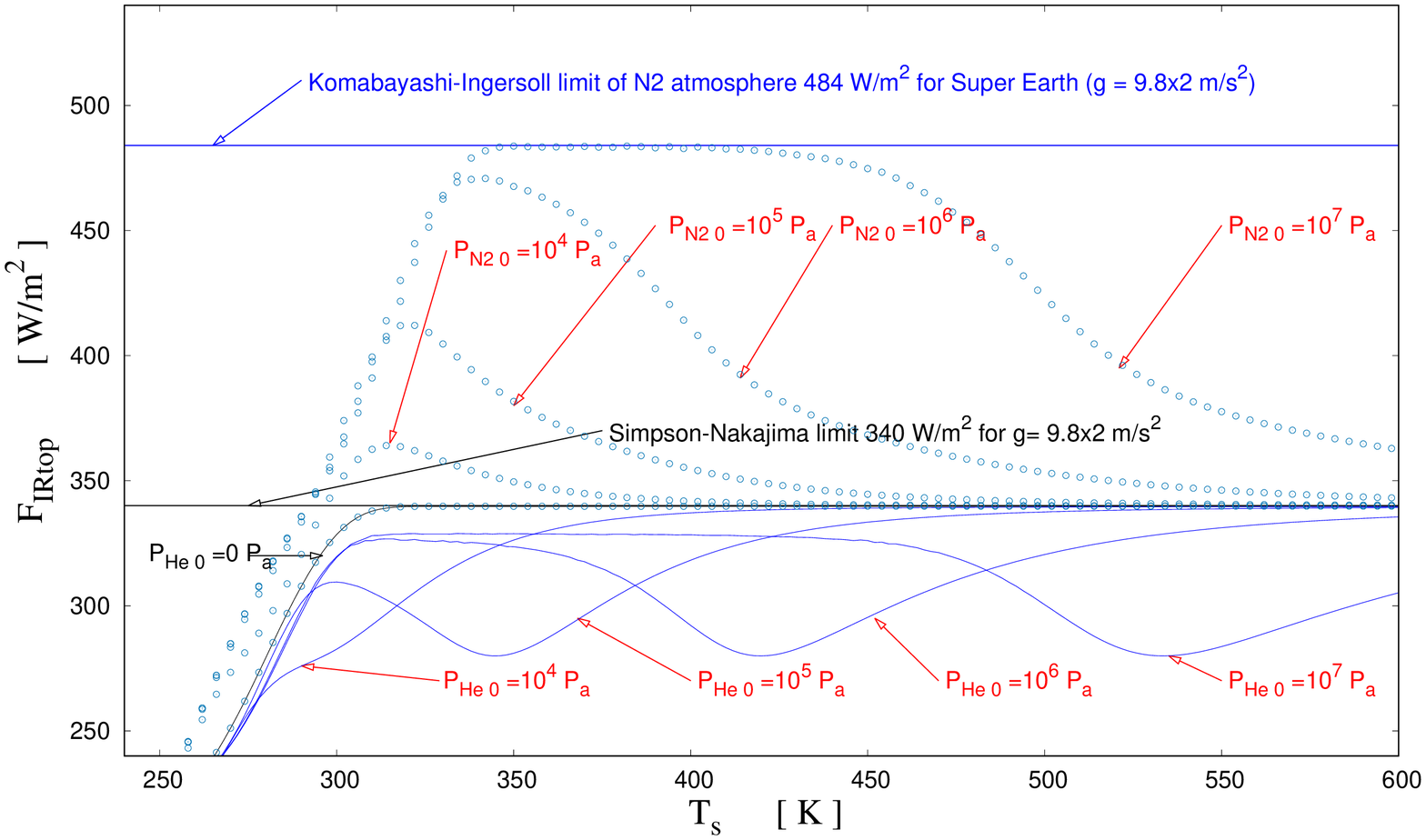}
\vspace{0.5cm}
\caption{For super-Earth case of background gases with He and N$_2$, the relationships between $T_s$ and $F_{IRtop}$ for several initial $p_{He\ 0}$ and $p_{N2\ 0}$ cases are presented.}
\end{center}
\end{figure}

\section{KI-limit dependence on the initial pressure}

It is noticed that the KI-limit depends on the initial pressure of the background component as well as $MW$ (molecular weight).

\begin{figure}[htbp]
\begin{center}
\includegraphics[keepaspectratio,scale=0.35]{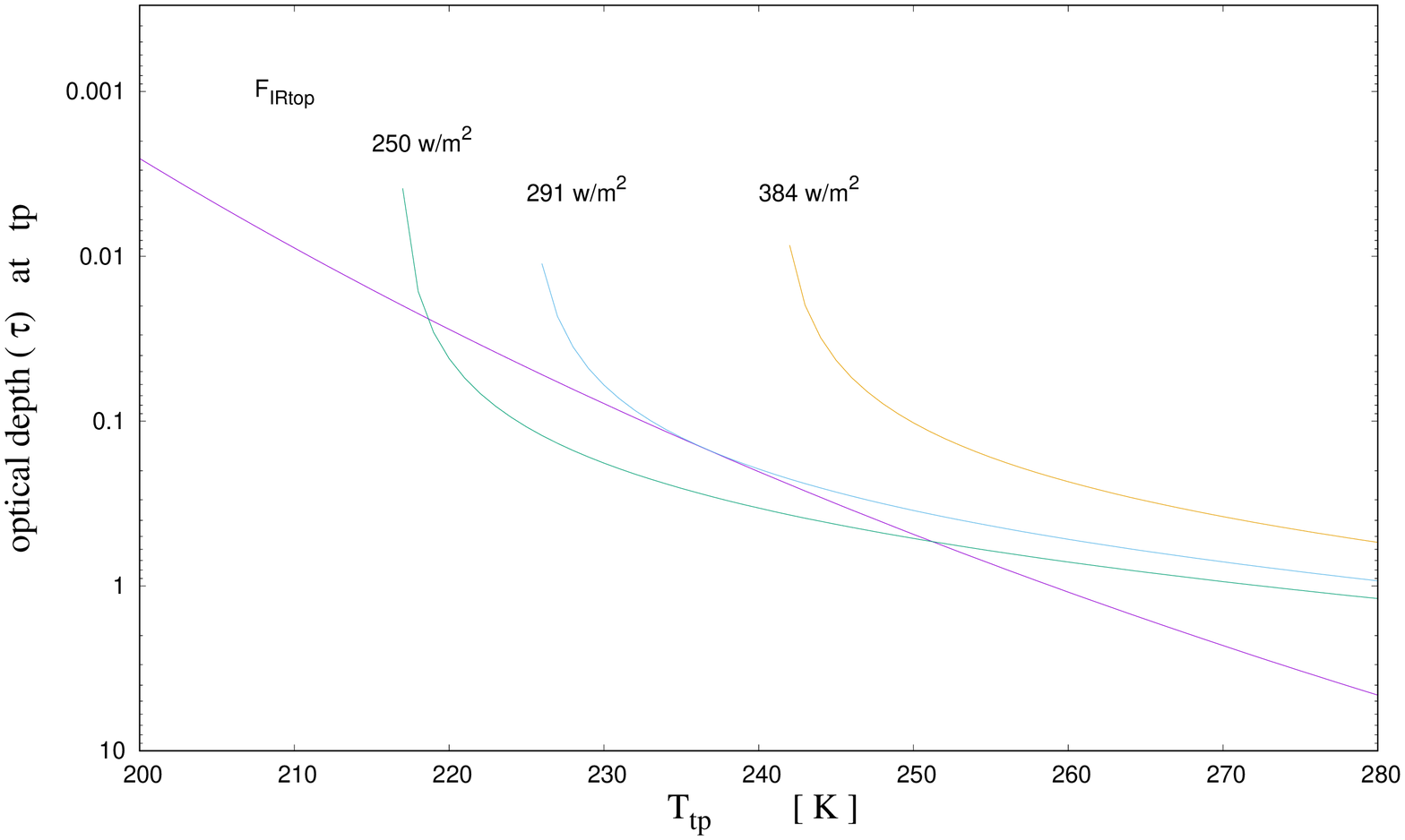}
\caption{The relationship between $\tau _{tp}$ and $T_{tp}$ of He atmosphere with $F_{IRtop}$ as a parameter for $p_{He\ 0}=10^6$ Pa.}
\end{center}
\end{figure}

In Fig. 23, the relationship between the optical depth $\tau _{tp}$ and the temperature $T_{tp}$ is presented of He atmosphere for $F_{IRtop}$ as a parameter with 250, 291, and 384 W/m$^2$ in $p_{He\ 0} = 10^6$ Pa.  The coloured curves represent Eq. (8) for $F_{IRtop}$ and the almost straight blue curve represents Eq. (9), where the suffix "$tp$" means the value at the tropopause (Nakajima  et al. 1992).

\begin{figure}[htbp]
\begin{center}
\includegraphics[keepaspectratio,scale=0.35]{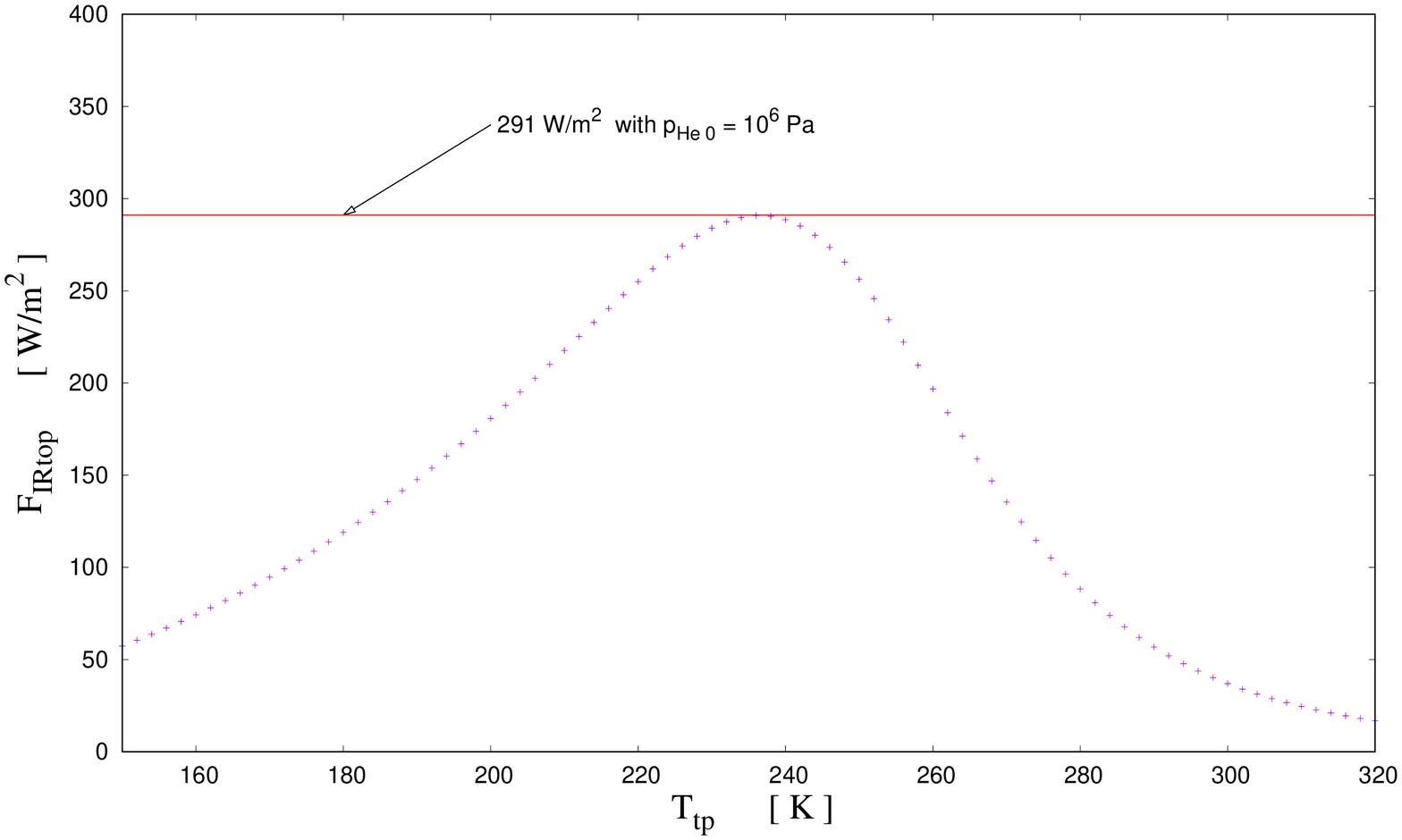}
\caption{The relationship between $T_{tp}$ and $F_{IRtop}$ for $p_{He\ 0} = 10^6$ Pa.  The KI-limit of this model for He atmosphere is $\simeq$ 291 W m$^{-2}$. }
\end{center}
\end{figure}

The relationship between $T_{tp}$ and $F_{IRtop}$ for $p_{He\ 0} = 10^6$ Pa is shown in Fig. 24.  The KI-limit of this model in He atmosphere is $\simeq$ 291 W m$^{-2}$ for $p_{He\ 0}\simeq 10^6$ Pa.

\begin{figure}[htbp]
\begin{center}
\includegraphics[keepaspectratio,scale=0.35]{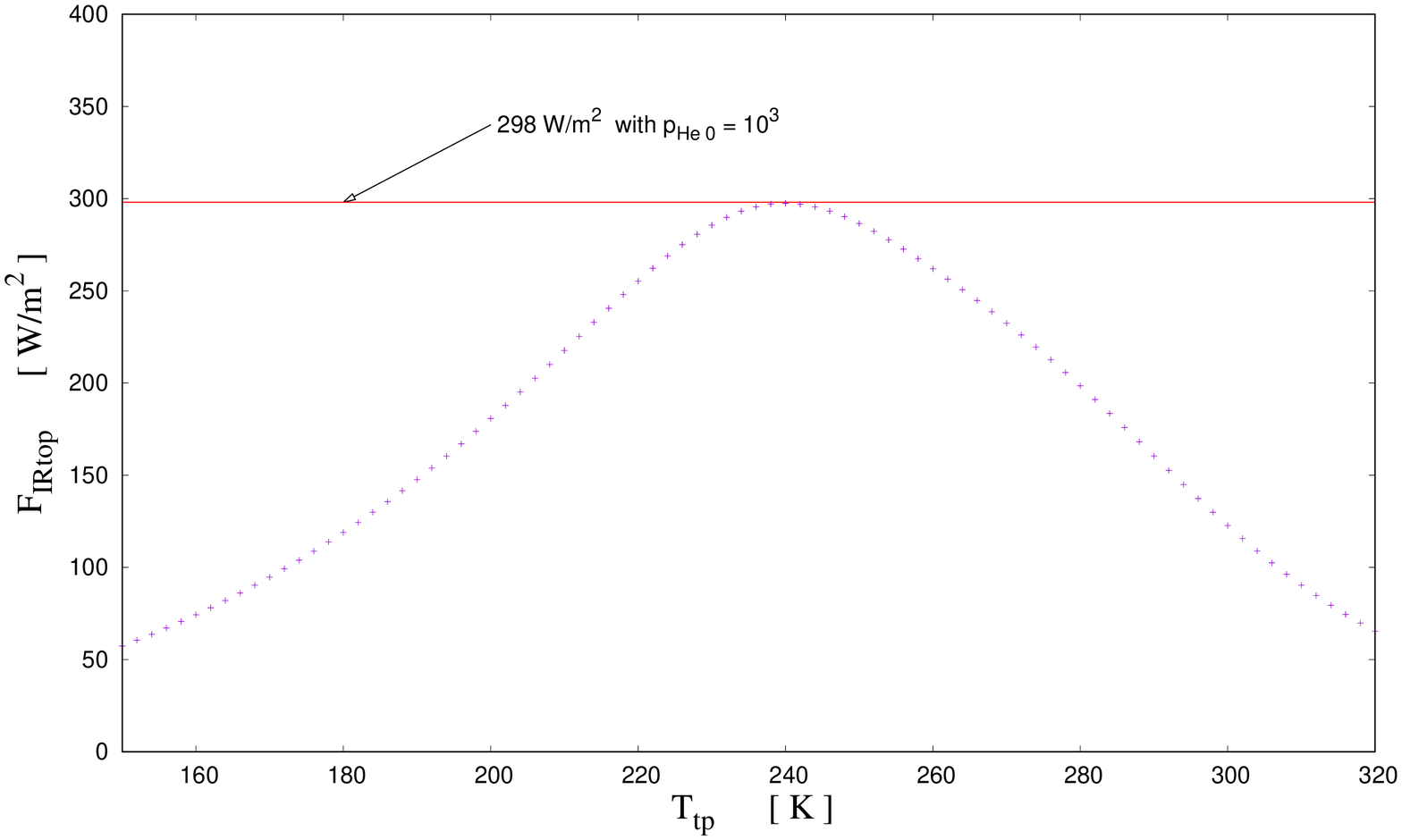}
\caption{The relationship between $T_{s}$ and $F_{IRtop}$ for $p_{He\ 0} \simeq 10^3$ Pa is presented.  The KI-limit of this model for He atmosphere is $\simeq$ 298 W m$^{-2}$, which is different from $\simeq$ 291 W m$^{-2}$ for $p_{He\ 0}\simeq 10^6$ Pa. } 
\end{center}
\end{figure}

In Fig. 25, the relationship between $T_{tp}$ and $F_{IRtop}$ for $p_{He\ 0} = 10^3$ Pa is presented where the KI-limit is $\simeq$ 298 W/m$^{2}$, which is different from $\simeq$ 291 W/m$^{2}$ for $p_{He\ 0} = 10^6$ Pa.  The KI-limit of He atmosphere is 291 $\sim$ 298 W/m$^{2}$ for $p_{He\ 0}\simeq 10^3 \sim 10^6$ Pa, which is shown in Fig. 26 as $MW=4$ case for He.  It shows that the KI-limit depends on the initial pressure $p_{He\ 0}$ of the background gas. 


\begin{figure}[htbp]
\begin{center}
\includegraphics[keepaspectratio,scale=0.35]{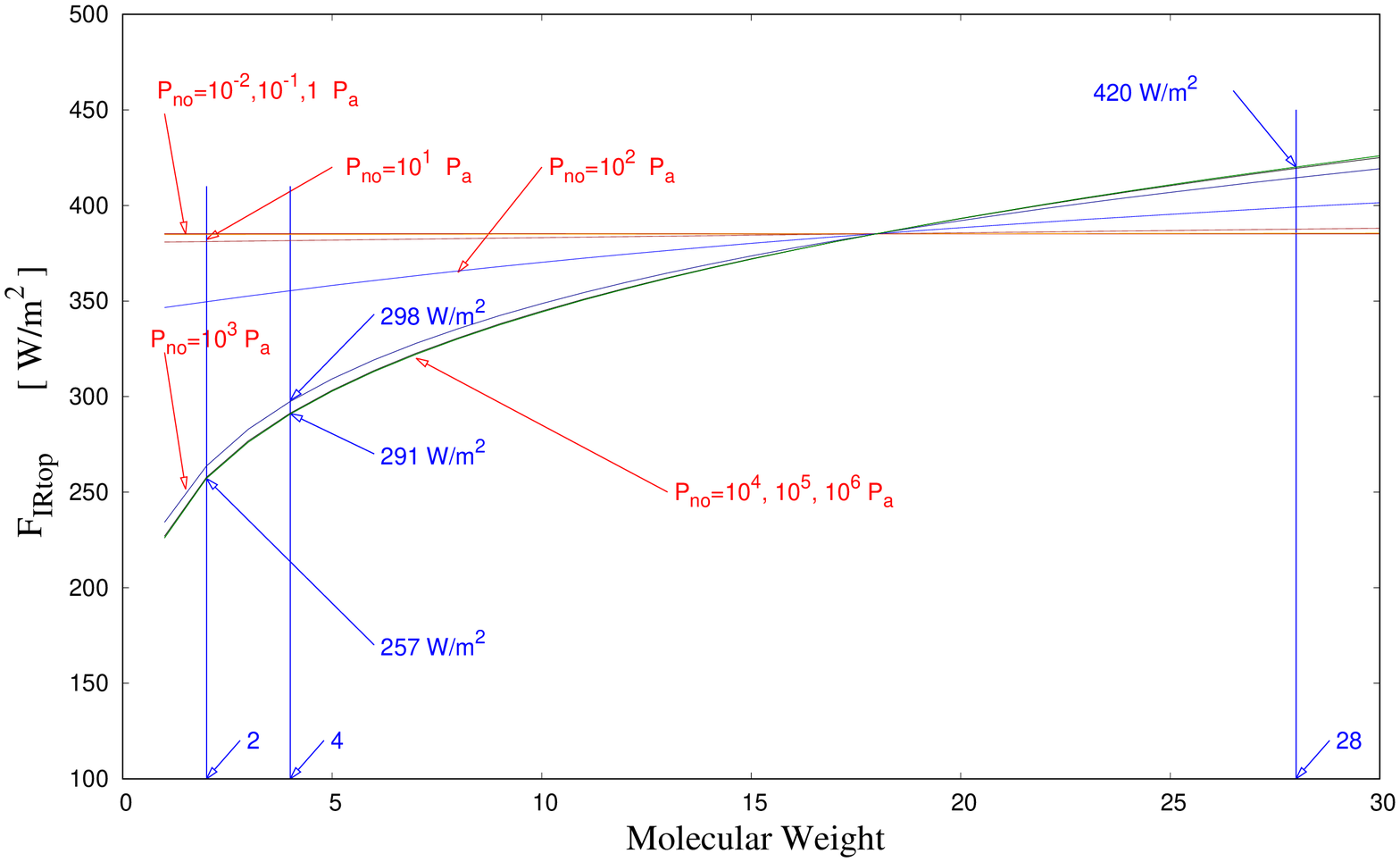}
\vspace*{0.5cm}
\caption{The KI-limit dependence on molecular weight for fixed initial pressure of the background gases.   The relationships between molecular weight and $F_{IRtop}$ for each initial pressure of the background gases is presented.  Molecular weight of H$_2$, He, and N$_2$ are shown as 2, 4, and 28, respectively.}
\end{center}
\end{figure}

In Fig. 26, it is calculated the KI-limit dependence on molecular weight for fixed initial pressure $p_{n\ 0}$ of the background gases.  The relationships between molecular weight and $F_{IRtop}$ for each initial pressure of the background gases is presented.  For example, in the case that the value of molecular weight is 18 where the molecular weight of every molecule is the same with water H$_2$O, $F_{IRtop}$ is the same for any initial pressure $p_{n\ 0}$.  On the other hand,  in the He background atmosphere the mean molecular weight (MMW) for $p_{He\ 0} \simeq 10^{-2}\sim 10^1$ Pa, the KI-limit is $ \simeq 390 $W/m$^2$ which is almost the same of $MW=18$, because the dominant component of the atmosphere is H$_2$O.  For $p_{He\ 0} \simeq 10^{4}\sim 10^6$ Pa, KI-limit of $MMW=4$ is $\simeq$ 291 W/m$^2$ where the dominant component of the atmosphere is He.  Even for fixed molecular weight of the background atmosphere, $MMW$ changes due to the increase of H$_2$O compared to the initial component pressure. The  KI-limit depends on the initial component pressure $p_{n\ 0}$.

About the saturation vapour pressure on temperature, we take Eq. (2), so the results in Fig. 26 seem to be independent on any peculiar temperature.


\section{Results and Discussion}


We have presented the results of the outgoing radiation for the different components of background gas under a simple model with gray approximation adopted by Nakajima et al. 1992. 
\vspace{0.4cm}
The main results of the paper are the following:

1.  It is studied the possibility of the various atmospheres over oceans.

2.  H$_2$ atmospheres as well as He are investigated.  To our concern,  the treating  

\hspace{0.4cm} He atmosphere in this style seems to be rather new.

3. The strange features are found for H$_2$ atmosphere as well as He that  $F_{IRtop}$ increases 

\hspace{0.4cm} to the first upper limit, then decreases and increase again to the second limit :

\hspace{0.4cm} named "Souffl$\acute{e}$ effect" by Koll \& Cronin (2019).

4. The first upper limit is the Komabayashi-Ingersoll limit (KI-limit), which is not stated 

\hspace{0.4cm} explicitly by Koll \& Cronin (2019).

5. The KI-limit value depends on the molecular weight and the initial pressure of 

\hspace{0.4cm} the background atmosphere (Sect. V).

6. The second limit is the Simpson-Nakajima limit (SN-limit) which is the atmosphere  

\hspace{0.4cm} mainly composed by vapour, called 'steam limit' (Koll \& Cronin 2019).

7. "Souffl$\acute{e}$ effect" is analysed by taking the UOD (Unit optical depth)  temperature 

\hspace{0.4cm} which has decreased and the UOD height which has increased.

8. H$_2$ atmosphere as well as He is the non-condensible background component where 

\hspace{0.4cm} the KI-limit is lower than the SN-limit.

9. The mixed gas of the relative realistic mixed atmosphere (H$_2 \simeq 72 \%  \ \&$ He $\simeq 28 \%)$

\hspace{0.4cm} is investigated.  The first upper limit is $\sim 267 $W/m$^2$ (Sect. VI).

10. The various approximate limits derived by Koll \& Cronin (2019) are investigated 

\hspace{0.4cm} and considered its applicability (Appendix B).

\vspace{0.4cm}
 From the above consideration, it becomes clear that it must be considered the atmospheric components of the extra terrestrial planets if one want to investigate "Habitable Zone".   
We would like to expect much more observations further to find out the life trace in extra-terrestrial planets (Hill et al. 2022).

\subsubsection{Injection flux vs. Distance from a parent star}

To apply the above results to the planet formation, it is better to consider the solar constant (1364 W/m$^2$) for Earth orbit where the gravitational acceleration as 9.8m/s$^2$.  Taking the spherical mean for the injected flux ($\times$ 1/4), the flux becomes 341 W/m$^2$.

 
 As the solar luminosity is almost 70$\%$ of present value at 4.6 Gy ago (Bahcall, Pinsonneault, \& Basu 2001), the flux is $\simeq$ 239 W/m$^2$.   If we take the albedo $\simeq$ 0.3 which is the corresponding value of Earth at present, the solar injection flux decreases to $\simeq$ 167 W/m$^2$.

The KI-limit for H$_2$ atmosphere is $\simeq$ 257 W/m$^2$ which is greater than the above $\simeq$ 239 W/m$^2$ and $\simeq$ 167 W/m$^2$ values.  Then it is stable for H$_2$ atmosphere for assumed planets in Earth orbit.  Even for super-Earth,  the KI-limit for H$_2$ atmosphere is $\simeq$ 295 W/m$^2$ which is greater than the above $\simeq$ 239 W/m$^2$and $\simeq$ 167 W/m$^2$ values.  Then it is stable for H$_2$ atmosphere for assumed super-Earth planets in Earth orbit.  

The KI-limit for He atmosphere is $\simeq$ 291 W/m$^2$ which is greater than the above $\simeq$ 239 W/m$^2$ and $\simeq$ 167 W/m$^2$ values.  Then it is stable for He atmosphere for the assumed planets in Earth orbit.  Even for super-Earth,  the KI-limit for He atmosphere is $\simeq$ 328 W/m$^2$ which is greater than the above $\simeq$ 239 W/m$^2$ and $\simeq$ 167 W/m$^2$ values.  Then it is stable for He atmosphere for the assumed super-Earth planets in Earth orbit.  

If the orbit has decreased to 0.5 Au, the injection flux increased four times as $\simeq$ 956 (=239$\times$ 4) W/m$^2$ and $\simeq$ 668 W/m$^2$.  Then  it becomes unstable for Earth and super-Earth planets, because the situations are over the KI-limit and SN-limit. They will be in a greenhouse runaway and/or moist greenhouse situation.

\begin{figure}[htbp]
\begin{center}
\includegraphics[keepaspectratio,scale=0.4]{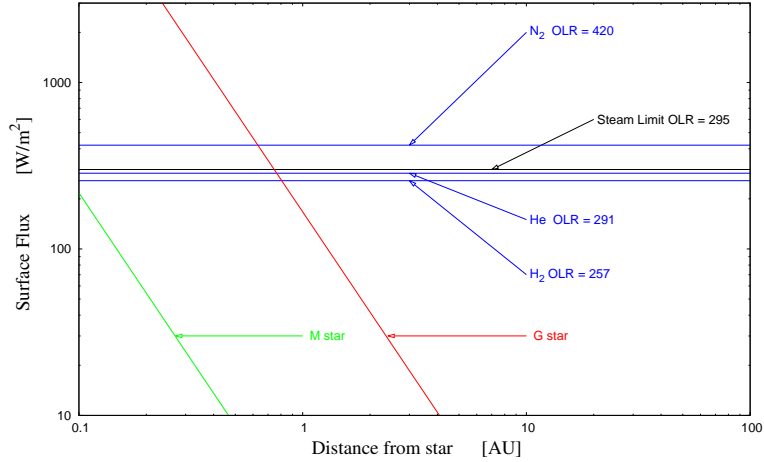}
\vspace{0.5cm}
\caption{Injection flux vs. distance from a parent star.}
\end{center}
\end{figure}

 In Fig. 27, the injection flux versus the distance from parent star is shown.   The star is taken for a G and an M host star by red and green line, respectively.  For an M-type star the luminosity is taken with 1.3 \% of the Sun (Pierrehumbert \& Gaidos 2011).  G star is normalized 167 W/m$^2$ at 1AU.  Three dilute limits for H$_2$ ($\simeq$ 257 W/m$^2$), He ($\simeq$ 291 W/m$^2$), and N$_2$ ($\simeq$ 420 W/m$^2$) atmospheres and the steam limit ($\simeq$ 295 W/m$^2$) are presented for reference.  The steam limit is shown by black.


  Under this model, it is found that the OLR upper limit decreases in the H$_2$ rich atmosphere for the decrease of the mean molecular weight.  Then it becomes clear that it is possible to become the runaway situation 
for super-Earth in the distance smaller than 1AU from the G type star (Sun).   Even if OLR has not reached the upper limit, there is a possibility to become dry up the ocean due to the high surface temperature of the greenhouse effect.

If there are extra planets with H$_2$ atmosphere over ocean and a life such as photosynthetic bacteria of the class Cyanobacteria (indicated by Pierrehumbert \& Gaidos, 2011), it must be an unstable situation for the chemical reaction between H$_2$ and O$_2$.   It could be a stable atmosphere with He and O$_2$.  
We would like to estimate the diffusion time of H$_2$ through the initial atmosphere mainly composed H$_2$ and He (Sekiya, Nakazawa \& Hayashi 1980; Wordworth 2012; Hu et al. 2015; Pahlevan et al. 2022).

\subsubsection{Mixed background gases with H$_2$ and He}
\begin{figure}[htbp]
\begin{center}
\includegraphics[keepaspectratio,scale=0.4]{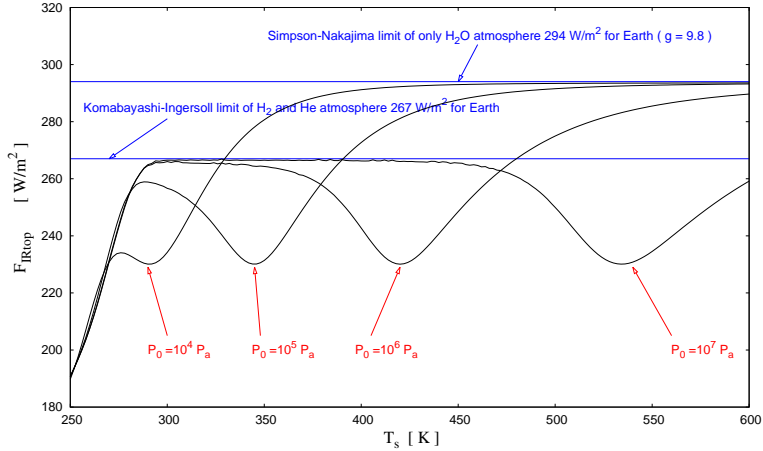}
\vspace{0.5cm}
\caption{It is investigated the gray calculations for mixed background gases with H$_2$ (weight X$\simeq 0.72$) and He (weight Y$\simeq 0.28)$ for $g=9.8 $m/s$^2$. }
\end{center}
\end{figure}

In Fig. 28, the gray calculations for mixed background gases with H$_2$ (weight X$\simeq 0.72$) and He (weight Y$\simeq 0.28)$ for $g=9.8 $m/s$^2$ are presented.  This mixed gas is considered to be the primordial gas from which Sun and Earth are formed.  The relationships between $T_s$ and $F_{IRtop}$ 
for several initial $p_{\ 0} (=10^4, 10^5, 10^6,$ and $10^7$ Pa) cases for Earth are shown.   The first upper limit for this atmosphere is $\sim$ 267 W/m$^2$), which has increased from H$_2$ of $\sim $ 257 W/m$^2$ (see Fig. 1).  The second limit (steam limit) is the same of H$_2$ and He as $\sim$ 294 W/m$^2$.


\subsubsection{Various uncertainty}

  There are various uncertainty factors to estimate the relationship between the surface temperature $T_{s}$ and 
outgoing infrared radiation at the top of the atmosphere $ F_{IRtop}$.  
Our results do not apply directly to any real planet history because of large uncertainties in our calculation due to the absence of clouds and the use of a one-dimensional model. 
 In order to determine quantitatively, it seems to be necessary to evaluate the parameters such as albedo, effects of clouds, and relative humidity, circulation of the atmosphere over the surface of the planet (Manabe \& Wetherald 1967, Zsom et al. 2013, Manabe \& Broccoli 2020). 
 It will be better to estimate the distribution of those parameters for the greenhouse effects and under the increase of solar luminosity. 

a. gray opacity  approximation

If one considers more than gray opacity approximation, one has to treat opacity, including the line by line treatments, Lorentz factor, pressure effect, random approximations, and further (Pierrehumbert 2010; Seager 2010), which make the physical understanding to be complicated.  

\appendix 
\renewcommand{\theequation}{A.\arabic{equation}} 
\setcounter{equation}{0} 
\section{Derivation of the equation (3) in relation to Pierrehumbert (2010)}
 In this appendix, it is explained that the equation of (4) in Nakajima et al. (1992) (Eq. (3) in this paper) is equivalent to the equation of (2.33) in Pierrehumbert (2010).  The notation is followed to Pierrehumbert (2010) and equation number is shown  as (Pie. 2.33).  As stated before, it is checked for the situation for the different molecular weights of condensible  and non-condensible  substances (Nakajima et al. have assumed the same for simplicity).

The equation in (Pie. 2.33) is written as

\begin{equation}
    \frac{d\  {\rm ln}\  T}{d\  {\rm ln}\  p_a} =\frac{R_a}{c_{pa}}\frac{1+\frac{L}{R_aT}r_{sat}}{1+\left(\frac{c_{pc}}{c_{pa}}+\left(\frac{L}{R_cT}-1\right)\frac{L}{c_{pa}T}\right) r_{sat}},
\end{equation}
where subscript $a$ and $c$ are related to non-condensible and condensible substance, respectively.  Other notations are the following.

A partial pressure of non-condensible substance is described as $p_a$ which is related to total pressure $p$ as $ p=p_a+ p_c$ where $p_c$ is the partial pressure of condensible substance (water vapour).
Then $dp=dp_a+dp_c$.  It is introduced saturation assumption that $p_c$ is replaced by $p_{c,sat}$.  Using Clausius-Clapeyron to re-write $dp_{c,sat}$ that
\begin{equation}
  \frac{dp_{c.sat}}{dT}=\frac{L}{R_cT^2}p_{c,sat},
\end{equation}
where $R_c$ is the gas constant for the substance which is condensing (water) as shown in (Pie. 2.25) and $L$ is the latent heat associated with the transportation to the more condensed phase.  

Then $R_ad \ {\rm ln}\ p_a$ of the second term in (Pie. 2.32) becomes as
\begin{equation}
R_ad\ {\rm ln}\ p_a=\frac{R_a}{p_a}(dp-\frac{Lp_{c,sat}dT}{R_cT^2}).
\end{equation}
So the second term in (Pie. 2.32) is written as
\begin{eqnarray*}
-(1+\frac{L}{R_aT}r_{sat})\frac{R_a}{p_a}(dp-\frac{Lp_{c,sat}}{R_cT^2}dT)\\
=-(1+\frac{L}{R_aT}r_{sat})\frac{R_a}{p_a}dp+(1+\frac{L}{R_aT}r_{sat})\frac{R_a}{p_a}\frac{Lp_{c,sat}}{R_cT^2}dT.
\end{eqnarray*}
The term of $dT$ is moved to the first term of (Pie. 2.32).

Then the equation of (Pie. 2.32) becomes for taking $\partial Q=0$ as 
\begin{equation}
\left[ {\tilde c}_{pa}+({\tilde c}_{pc}+(\frac{L}{R_cT}-1)\frac{L}{T})r_{sat}+(1+\frac{L}{R_aT}r_{sat})\frac{R_a L p_{c,sat}}{p_aR_cT} \right]  d\ {\rm ln}\ T
\end{equation}
\begin{equation}
=\left [  (1+\frac{L}{R_aT}r_{sat})\frac{R_ap}{p_a}  \right]  d\ {\rm ln}\ p.
\end{equation}

Then taking $r_{sat}=\epsilon p_{c,sat}/p_a,  \ \epsilon=M_c/M_a, \  p_a=\rho _aRT/M_a,  \ p_c=p_{c,sat}=\rho _cRT/M_c, \  x_n=p_n/p=p_a/p,$ and $x_c=p_c/p,  \ R_a=R/M_a, \  R_c=R/M_c, \ {\it l}=LM_c, \ c_{pa}={\tilde c}_{pa} M_a,  \ c_{pc}={\tilde c}_{pc} M_c, $ where $M_a$ and $M_c$ are molecular weight of the substanc a (=n: non-condensible  background substance, eg. H$_2$, He, N$_2$, and air: suffix 'a' comes  from Pierrehumbert and 'n' comes from Nakajima et al.), and c (=v: condensible  substance, eg. water vapour), the above equations have changed to the followings.

Using $(R_a L p_c)/(p_a R_c T)=(M_c p_c L)/(M_a p_a T)=(r_{sat} L)/T$, the square bracket of Eq. (A. 4) becomes as
\begin{eqnarray*}
\left[\cdots\right]&=&\left[ {\tilde c}_{pa}+({\tilde c}_{pc}+(\frac{L}{R_cT}-1)\frac{L}{T})r_{sat}
+(1+\frac{Lr_{sat}}{R_aT})\frac{LR_a p_{c,sat}}{p_aR_cT} \right]\\
&=&\left[ {\tilde c}_{pa}+({\tilde c}_{pc}+\frac{L}{R_cT}\frac{L}{T})r_{sat}-\frac{L}{T}r_{sat}
+r_{sat}\frac{L}{T}+\frac{Lr_{sat}}{R_aT}\frac{r_{sat} L}{T} \right] \\
&=&\left[ {\tilde c}_{pa}+{\tilde c}_{pc}r_{sat}+\frac{l^2}{M_c^2}\frac{M_c}{R}\frac{1}{T^2}\frac{M_c}{M_a}\frac{p_c}{p_a}
+\frac{l^2M_a}{M_c^2RT^2}\frac{M_c^2}{M_a^2}\frac{p_c^2}{p_a^2}\right]\\
&=&\frac{1}{M_a}\left[ c_{pa}+c_{pc}\frac{x_v}{x_n}+\frac{l^2}{RT^2}\frac{x_v}{x_n}
+\frac{l^2}{RT^2}\frac{x_v^2}{x_n^2}\right]\\
&=&\frac{c_{pa}}{M_a x_n}\left[x_n+x_v\frac{c_{pc}}{c_{pa}}+\frac{x_v}{x_n}\frac{l^2}{RT^2c_{pa}}\right] ,
\end{eqnarray*}
where we use $x_n+x_c=1$.

The square bracket of Eq. (A. 5) becomes as
\begin{eqnarray*}
\left[\cdots\right]&=&\frac{R_a}{x_n}\left [  (1+\frac{Lr_{sat}}{R_aT}) \right] \\
&=&\frac{R_a}{x_n}\left [1+\frac{l}{RT}\frac{x_c}{x_n} \right].
\end{eqnarray*}
Considering $d\ {\rm ln}\ T/d\ {\rm ln}\ p=\frac{p}{T}\frac{\partial T}{\partial p}$, the equation (Pie. 2.33) becomes as

\begin{equation}
    \left(\frac{\partial T}{\partial p} \right)=\frac{\frac{RT}{pc_{pn}}+\frac{x_v^*}{x_n}\frac{l}{pc_{pn}}}{x_n+x_v^*\frac{c_{pc}}{c_{pn}}+\frac{x_v^*}{x_n}\frac{l^2}{RT^2c_{pn}}},
\end{equation}
where we neglect the differences between partial derivative and total derivative, and take $x_c=x_v^*$ and $c_{pa}=c_{pn}$.  \cite{When}


Then the derivation of equation (4) in Nakajima et al. is equivalent to the equation of (Pie. 2. 33).

It is also equivalent to the equations of (3.7) and (3.14) in Houghton (1977)

\begin{equation}
  dp= -g\rho_a(1+\xi)dz 
\end{equation}
and 
\begin{equation}
-\frac{dT}{dz}=\frac{g}{{\tilde c}_{pn}}\frac{(1+Lp_{c,sat}M_c/pRT)(1+(p_{c,sat}\epsilon/p)}{1+(\epsilon p_{c,sat}/p{\tilde c}_{pn})(\tilde c_{pc}+(dL/dT)-L/T)+(\epsilon p_{c,sat}L^2M_c/{\tilde c}_{pn}pRT^2)}.
\end{equation}

If the term $ dL/dT- L/T$ in the denominator is neglected for its small values, the above two equations becomes equivalent to the equation (3) in this paper and equation of (4) in Nakajima et al. (1992).


\renewcommand{\theequation}{B.\arabic{equation}} 
\setcounter{equation}{0} 
\section{Trial to derive the limit values}

\subsection{MMW matters (Koll \& Cronin 2019)}
First, taking the water vapour scale height $H_{v}, \rho _{v}=\rho _{v , 0} e^{-{z/H_{v}}}$, the optical depth is

\begin{equation}
  \tau=\int _0^{\infty} \kappa _{v} \rho _{v} dz=\kappa _{v} \rho _{v ,0} \int _0^{\infty} e^{-z/H_{v}}dz=\frac{H_{v}}{l_0},
\end{equation} 
where $l_0=1/(\kappa _{v} \rho _{v ,0}) $ (in general, the subscript $v$ will denote quantities related to water vapour).

Using the ideal gas law for water vapour, $\rho _{v}=e^*/(R_{v}T)$ where $e^*$ is the saturation vapour pressure, and
 the Clausius-Clapeyron relation, $d$ ln $e^* / d$ ln $T=L_{v}/(R_{v}T)$ where $L_{v}$ is the latent heat of vapourization:
\begin{equation}
  (-H_{v})^{-1}=\frac{1}{\rho_{v}}\frac{d \rho _{v}}{dz} =\frac{1}{e^*}\frac{d e^*}{dz}-\frac{1}{T}\frac{dT}{dz}=\left(\frac{1}{e^*}\frac{d e^*}{dT}-\frac{1}{T} \right) \frac{dT}{dz},
\end{equation} 
and 
\begin{equation}
 =\left(\frac{L_{v}}{R_{v}T}-1\right) \frac{1}{T}\frac{dT}{dz} \simeq  \frac{L_{v}}{R_{v}T} \frac{1}{T}\frac{dT}{dz}=\frac{L_{v}}{R_{v}T} {H_T}^{-1},
\end{equation} 
where $L_{v}/R_{v}T\gg 1 $ is used.

Here it is denoted $H_T=(-1/T\times dT/dz)^{-1}$ as the temperature scale height, which is related to the water vapour scale height as
\begin{equation} 
  H_{v}=\frac{R_{v}T}{L_{v}}H_T,
\end{equation} 
then it is accepted $H_T \gg H_{v}$ for water vapour and a wide range of other condensible gases.

In the low temperature limit, the lapse rate is assumed to be the dry adiabat, $dT/dz=-g/c_p$, and
\begin{equation} 
  H_T^{dilute}=\frac{c_p T}{g}.
\end{equation} 

At high temperature the total pressure is dominated by the saturation vapour pressure of water and the temperature scale height is given by

\begin{equation} 
  H_T^{steam}=-T\frac{dz}{dT}=-T\frac{dz}{de^*}\frac{de^*}{dT}=T\frac{1}{\rho_{v} g}\frac{L_{v}e^*}{R_{v}T^2}=\frac{L_{v}}{g},
\end{equation} 
where  the hydrostatic equation, $de^*/dz=-\rho_{v}g$, is used.

It should be noted that at a temperature of 400 K with an Earth-like gravity, $H_T^{dilute} \simeq 350$ km in H$_2$, whereas $H_T^{dilute} \simeq 40$ km in N$_2$. The main difference is due to the low molecular weight of H$_2$ relative to N$_2$ (see Fig. 5).

\subsection{Trial to derive the steam and dilute limit values}

Koll \& Cronin (2019) have tried to derive an approximate analytical expression for the OLR limit values of the steam limit (SN-limit)  and dilute limit (KI-limit).  They have approximated the saturation pressure with temperature as

\begin{equation}
  p^*(T)=p^*_0\left( \frac{T}{T_0} \right)^{\gamma},
\end{equation} 
 taking the Clausius-Clapeyron relation d ln $p^*/$d ln $T \simeq L_{v}/(R_{v}T) = \gamma $.  They have expressed the relation between optical depth and temperature  as a power law (see Koll \& Cronin 2019)

 \begin{equation}
  \tau(T)=\kappa H_T \frac{p^*_0}{L_{v}}\times \left( \frac{T}{T_0} \right)^{\gamma},
\end{equation} 
where $T_0$ is a reference temperature close to the interest range.

OLR is said to be equal to
 \begin{equation}
  {\rm OLR}=\sigma T_s^4e^{-\tau}+ \int ^{\tau}_{0} \sigma T(\tau ')^4 e^{-\tau '} d\tau ',
\end{equation} 
the first term can be neglected as optical depth increasing and the upper limit of the integral is replaced with infinity

\begin{eqnarray}           
{\rm OLR}_{\infty} \simeq \int _{0}^{\infty} \sigma T(\tau )^4 e^{-\tau } d\tau ,\\
          =\sigma {T_0}^4 \int _{0}^{\infty} \left( \frac{\tau}{\tau _0} \right) ^{4/\gamma} e^{-\tau} d\tau ,\\
          =\sigma {T_0}^4 {\tau_0}^{-4/\gamma} \int _{0}^{\infty} \tau ^{4/\gamma} e^{-\tau } d\tau ,\\
         {\rm OLR}_{\infty} \simeq \sigma {T_0}^4 \times \frac{\Gamma (1+4/\gamma)}{\tau _0^{4/\gamma}}.
\end{eqnarray} 
  
Here $\Gamma$ is the gamma function, defined by $\Gamma (s)=\int _0^{\infty}x^{s-1} $exp$(-x) d x$, and $\tau_0 $ denotes for steam limit 
\begin{equation}
  \tau _0 = \frac {\kappa _{v} p^*(T_0)}{g}, 
\end{equation} 
and for dilute limit
\begin{equation}
  \tau _0  = \frac {\kappa _{v} p^*(T_0)}{g} \frac{c_p T_0}{L_{v}},
\end{equation} 
where it is taken $c_p=7R/2$ for H$_2$ and N$_2$ diatomic molecule and $c_p=5R/2$ for He monoatomic molecule.  The latent heat $L_v$ is tentatively taken, for the moment, as $2265-4.186\times (T_0-373)$ (J/g), where 2265 (J/$g$) is the vapourization energy per gram at 373K and 4.186 (J/($g\cdotp$ K)) is the specific heat of water, approximately.

\begin{figure}[htbp]
\begin{center}
\includegraphics[keepaspectratio,scale=0.4]{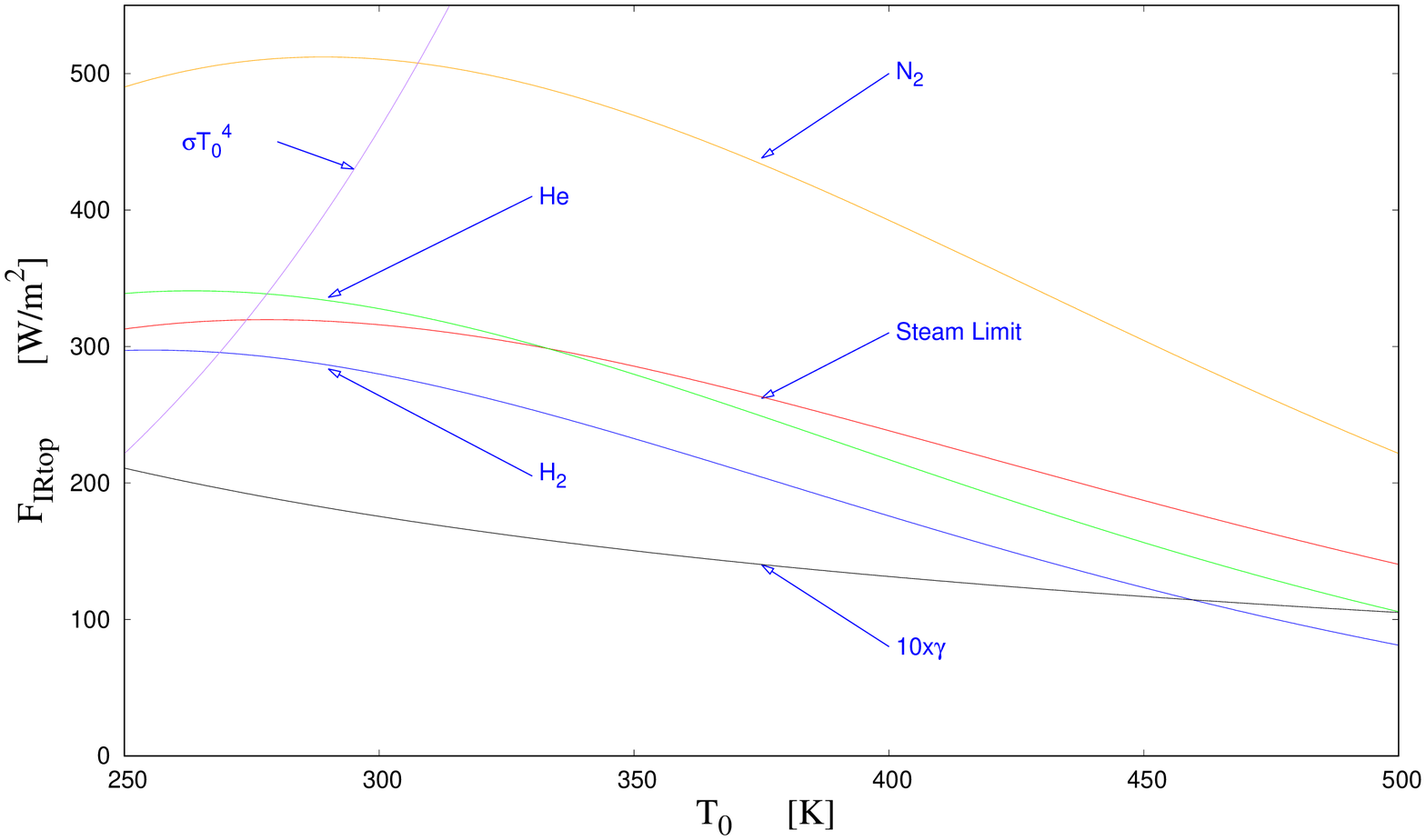}
\vspace{0.5cm}
\caption{Dilute limit dependence on $T_0$ for H$_2$, He, and N$_2$ atmospheres given by Eqs. (B.13) and (B.15) are presented.}
\end{center}
\end{figure}

Dilute limit dependence on $T_0$ for H$_2$, He, and N$_2$ atmospheres given by Eqs. (B.13) and (B.15) are presented by blue, green, and orange curve in Fig. 29, respectively. 
For the steam limit given by Eqs. (B.13) and (B,14) is presented by red curve in Fig. 29.  
They are almost constant or monotonically decreasing with $T_0$, however the value of $\sigma T_0 ^4$ has changed which is shown for reference by purple curve in Fig. 29.

Taking $T_1=T_0-\Delta T$ and $T_2=T_0+\Delta T$, $\gamma $ for $T_0$ is taken as a mean value of $\gamma _i$ as $(\gamma _1 + \gamma _2)/2$ where
\begin{equation}
  \gamma_i = \frac{L}{R} \times \left ( \frac{1/T_i-1/T_0}{{\it ln}T_0-{\it ln} T_i}\right),
\end{equation}  
being i (=1, 2) and $\Delta=25$.  Ten times of $\gamma$ is presented in Fig. 29 in black curve as $ 10\times \gamma$ and some $\gamma$ values are shown in Table I.

It must be noticed that He dilute limit (green line) lies below the steam limit (red line) beyond $T_0 \simeq 330$ K, which is not indicated by Koll \& Cronin 2019.

For taking $T_0 \simeq 300 $ K where $ \gamma \simeq 17.55$,  OLR $\simeq 316$ W/m$^2$ for steam limit (calculated result is $\simeq 294$ W/m$^2$ (Fig. 1); then 316/294$\simeq $1.075: the difference is $\simeq$ 7.5\%).  The dilute limit of H$_2$ atmosphere is $\simeq 280$ W/m$^2$ (calculated result is $\simeq 257$ W/m$^2$ (Fig. 1), then 280/257 $\simeq$ 1.089: the difference is $\simeq$ 8.9 \%).  The dilute limit of He atmosphere is $\simeq 328$ W/m$^2$ (calculated result is $\simeq$ 291 W/m$^2$ (Fig. 12), then 328/291 $\simeq$ 1.127: the difference is $\simeq$ 13 \%).  The dilute limit of N$_2$ atmosphere is $\simeq 511$ W/m$^2$ (calculated result is $\simeq 420$ W/m$^2$ (Fig. 9), then 511/420 $\simeq$ 1.217: the difference is $\simeq$ 22 \%). 

 Although they have shown some limiting values, it seems to be a little bit difficult to accept that the limiting values could approximately represent the calculated values. The absolute limiting values seems to have some differences (dilute limit of N$_2$ differs about 22 \%) for $T_0 \simeq$ 300K and (dilute limit of H$_2$ is only about 32 \% (differs 68 \%) for $T_0 \simeq$ 500K. They are shown in Fig. 29 and particular values are presented in Table I.  

Let us consider the limiting values from the different point.  The relative differences of the limiting values could be understood by the above Eq. (B.14).   


1.  The difference of the accelerating gravity: if $g$ has changed to $\alpha g$ by factor $\alpha$, $\tau_0$ has changed to $\tau_0/\alpha$ and OLR$_{\infty}$ has changed by factor $\alpha ^{4/\gamma}$.  Taking $\alpha =2$ and $\gamma \simeq 17.55$, it becomes $2^{0.2279} \simeq 1.1711$.  The steam limit has changed from $\simeq$ 294 W/m$^2$ (Fig. 1) to $\simeq$ 340 W/m$^2$ (Fig. 10) which is almost the same as $340/294 \simeq 1.1564  (1.1711/1.1564 \simeq 1.0128$).  The difference is within 1.3\%.
. 
The dilute limit for H$_2$, it has changed from $\simeq$ 257 W/m$^2$ (Fig. 1) to $\simeq$ 295 W/m$^2$ (Fig. 10) where 295/257 $\simeq $1.1479 (1.1711/1.1479 $ \simeq $1.0202).  The difference is within 2.1 \%.  The dilute limit for He, it has changed from $\simeq$ 291 W/m$^2$ (Fig. 12) to $\simeq$ 328 W/m$^2$ (Fig. 21) where 328/291 $\simeq$ 1.12714 (1.1711/1.2714 $\simeq$ 0.9211). The difference is within 7.9 \%.  The dilute limit for N$_2$, it has changed from $\simeq$ 420 W/m$^2$  (Fig. 9) to $\simeq$ 484 W/m$^2$ (Fig. 22) where 480/420 $\simeq$ 1.14286 (1.1711/1.14286 $\simeq$ 1.0247). The difference is within 2.5 \%.

2. The difference of the mean molecular weight: $c_p$ is the specific heat per gram related to the specific heat per mol $ \hat {c}_p=c_p \times MW$ where $MW$ is the molecular weight.  Then if the molecular weight has increased by factor $\beta$,  $\tau_0$ has decreased by factor $\beta$.  The $\tau_0$ has changed to $\tau_0/\beta$ and OLR$_{\infty}$ has changed by factor $\beta ^{4/\gamma}$.  Taking $\beta =2$ from H$_2$ to He and $\gamma \simeq 17.55$, it becomes $2^{0.22792} \simeq 1.1711$.  The dilute limit has changed from $\simeq$ 257 W/m$^2$ (Fig. 1) to $\simeq$ 291 W/m$^2$ (Fig. 12) which is almost the same as 291/257  $\simeq 1.132(1.17114/1.1323  \simeq 1.0343$).  The difference is within 3.5 \%.

Taking $\beta =14$ from H$_2$ to N$_2$ and $\gamma \simeq 17.55$, it becomes $14^{0.22792} \simeq 1.8248$.  The dilute limit has changed from $\simeq$ 257 W/m$^2$ (Fig. 1) to $\simeq$ 420 W/m$^2$ (Fig. 9) which is almost the same as 420/257 $\simeq 1.63424 (1.8248/1.63424  \simeq 1.117$) .  The difference is within 12 \%.

Taking $\beta =7$ from He to N$_2$ and $\gamma \simeq 17.55$, it becomes $7^{0.222792} \simeq 1.5582$.  The dilute limit has changed from $\simeq$ 291 W/m$^2$ (Fig. 12) to $\simeq$ 420 W/m$^2$ ( Fig. 9)which is almost the same as 420/291 $ \simeq 1.4433 (1.5582/1.4433\simeq 1.0796).$  The difference is within 8.0 \%.

 It is a rather good  approximation to estimate the OLR$_{\infty}$ by the variation of gravity (differences are within 7.9 \%) and molecular weight (differences are within 12 \%).


\vspace{0.1cm}
\begin{table}[htbp]
\caption{Values of $\gamma$, the steam limit (SL), dilute limit of H$_2$, He and N$_2$ for T$_0$ are presented.
 R1, R2, R3 and R4 show the ratio of the approximate value to the calculated value. }
\vspace{0.3cm}
\begin{tabular}{|c||r|r|r|r|r|r|r|r|r|}  \hline
\ $T_0$ K  \ & $\gamma$ \  \  \ &  \ \ SL \ & \ \  \ R1 \ \  &  \ \  \ H$_2 \ $ & \ \ \  \ R2 \ \ &  \  \ He \  & \ \  \ R3 \ \  & \  \  \ N$_2 \ $ \ & \  \ R4  \ \ \\ \hline \hline
 300 & 17.55 & 316 & 1.078 & 280 & 1.089 & 328 &1.126 &  511 & 1.216 \\ \hline
 350 & 15.03 & 286 & 0.975 & 233 & 0.905 & 280 & 0.961 & 469 & 1.117 \\ \hline
 400 & 13.15 & 238 & 0.813 & 176 & 0.684 & 217 & 0.746 & 393 & 0.935 \\ \hline
 450 & 11.68 & 187 & 0.639 & 123 & 0.480 & 156 & 0.537 & 304 & 0.725 \\ \hline
 500 & 10.51  &  140 & 0.479 &  81.2  &  0.316 & 106 & 0.363 & 222 & 0.528 \\ \hline
\end{tabular}
\end{table}

\vspace{0.1cm}

In the Table I the particular values of $\gamma$, the steam limit and the dilute limit for of H$_2$, He and N$_2$ are shown, where SL, H$_2$, He and N$_2$ represent steam limit, dilute limit of H$_2$, He and N$_2$, respectively.
 R1, R2, R3 and R4 show the ratio of the approximate value to the calculated value, as SL/293, H$_2$/257, He/ 291 and N$_2$/420, respectively.  For example in $T_0 \simeq 300$ K, the difference of the steam limit is $\simeq$ 7.8 \%.  The difference of the dilute limit of H$_2$ is  $\simeq$ 8.9 \%.  If $T_0$ has changed to 500 K, the differences have increased much further. 

From the values in the Table I, the approximate analytical expressions seems to be applicable in the temperature $T_0 \simeq 300 \sim 350$ K  where the differences are within and around 20 \% (not bad).  For $T_0 \geq 400$ K, the differences are worse than 20 \%.  The reason  is not so clear for the moment.




\renewcommand{\theequation}{C.\arabic{equation}} 
\setcounter{equation}{0} 
\section{Try to derive Eq. (33) in Koll \& Cronin (2019)}

We have tried to derive the following three equations in  Koll \& Cronin (2019), 
\begin{eqnarray*}
  \frac{dp}{dz} {\vert} _{steam} & =& -\bar{\rho} g,   \hspace{8.8cm}  ({\rm KC} \ 32)   \\
                                        & \approx & \rho_{\nu} g \times \left( 1+\frac{R_d-R_{\nu}}{R_d} \frac{p_d}{e^*} \right),  \hspace{5cm}  ({\rm KC} \ 33)    \\
H_T &=& \frac {L_{\nu}}{g} \left( 1+ \frac{R_d-R_{\nu}}{R_d} \frac{p_d}{e^*} \right),   \hspace{5.5cm}     ({\rm KC} \ 34)       
\end{eqnarray*}

where (KC 32) represents Eq. (32) in Koll \& Cronin (2019) and so on.

Using $p=p_{\nu}+p_d$=$e^* +p_d$ and $ x_d=p_d/p$, the equation of  (KC 32) becomes
\begin{eqnarray*}
\frac{dp}{dz} {\vert} _{steam} & =&  \frac{dp}{dz}-\frac{dp_d}{dz} \approx -(\rho _{\nu} +\rho _d)g+(\rho _{\nu} +\rho _d)x_d g  \\
& =&-\rho_{\nu}g(1-x_d) \left(1+\frac{\rho_d}{\rho _{\nu}}\right)= -\rho_{\nu}g \left(1-\frac{1}{1+p_d/e^*}\frac{p_d}{e^*}\right) \left(1+\frac{p_d}{e^*}\frac{R_{\nu}}{R_d}\right)   \\
  & \approx  & -\rho_{\nu}g \left(1+\left(\frac{R_{\nu}}{R_d}-\frac{1}{1+p_d/e^*}\right)\frac{p_d}{e^*}\right)     \\
  & \approx & -\rho_{\nu}g \left(1+\left(\frac{R_{\nu}-R_d}{R_d} \right) \frac{p_d}{e^*} \right),
\end{eqnarray*}
where we use the relation $p_{\nu}=e^*=\rho_{\nu} R_{\nu}T, p_d=e^*=\rho_d R_dT$, and $p_d/e^* \ll 1$.  It must be noticed that signatures in and out of the parentheses
 are different from (KC 33).  We believe that those are typos.

Using the Clausius-Clapeyron relation, $d \, {\rm ln} \, e^*/d \, {\rm ln} \,  T = L_{\nu}/(R_{\nu}T) $, Eq. (KC 34) becomes as
\begin{eqnarray*}
 H_T&=&-T\frac{dz}{dp}\frac{dp}{dT} \approx -T\frac{dz}{de^*}\frac{de^*}{dT} \approx -\frac{T}{\rho_{\nu} g}\left(1+\frac{R_{\nu}-R_d}{R_d}\frac{p_d}{e^*} \right) ^{(-1)}\frac{L_{\nu}e^*}{R_{\nu} T^2}   \\
  & \approx & \frac{L_{\nu}}{g} \left( 1-\frac{R_{\nu}-R_d}{R_d} \frac{p_d}{e^*} \right) = \frac{L_{\nu}}{g} \left( 1+\frac{R_d-R_{\nu}}{R_d}\frac{p_d}{e^*} \right) . 
\end{eqnarray*}
where it must be noticed that the signature of the term in the parenthsis is exact of Eq. (KC 34).

 \end{document}